\documentclass[prb,showpacs,floatfix,preprint,amsmath,amssymb]{revtex4}

\usepackage{graphicx}
\usepackage{dcolumn}
\usepackage{bm}
\usepackage{epsfig}
\DeclareMathAlphabet{\mathpzc}{OT1}{pzc}{m}{it}

\newcommand{\dummy}

\begin{document}

\tighten

\title{Static and Dynamic Critical Behavior of a Symmetrical Binary
Fluid: A Computer Simulation}

\author{Subir K. Das,$^{\text{1}}$ J\"urgen Horbach,$^{\text{2}}$ Kurt Binder,$^{\text{2}}$
Michael E. Fisher,$^{\text{1}}$ and Jan V. Sengers$^{\text{1}}$} 
\affiliation{$^1$ Institute for Physical Science and Technology, University of
Maryland, College Park, MD 20742, USA \\
$^2$ Institut f\"ur Physik, Johannes Gutenberg Universit\"at
Mainz,
Staudinger Weg 7, 55099 Mainz, Germany}
\date{\today}

\begin{abstract}

A symmetrical binary, A+B Lennard-Jones mixture is
studied by a combination of semi-grandcanonical Monte Carlo
(SGMC) and Molecular Dynamics (MD) methods near a liquid-liquid critical
temperature $T_c$. Choosing equal chemical potentials for the two 
species, the SGMC switches identities (${\rm A} \rightarrow {\rm B} \rightarrow {\rm A}$) 
to generate well-equilibrated
configurations of the system on the coexistence curve
for $T<T_c$ and at the critical concentration, $x_c=1/2$, for
$T>T_c$. 
A finite-size scaling analysis of the concentration
susceptibility above $T_c$ and of the order parameter below
$T_c$ is performed, varying the number of particles from
$N=400$ to $12800$. The data are fully compatible with the expected
critical exponents of the three-dimensional Ising 
universality class.

The equilibrium configurations from the SGMC runs are used as
initial states for microcanonical MD runs, from which transport
coefficients are extracted. Self-diffusion coefficients are
obtained from the Einstein relation, while the interdiffusion
coefficient and the shear viscosity are estimated from
Green-Kubo expressions. As expected, the self-diffusion constant does
not display a detectable critical anomaly. With appropriate
finite-size scaling analysis, we show that the simulation data
for the shear viscosity and the mutual diffusion constant are quite consistent
both with the theoretically predicted behavior, including the critical exponents 
and amplitudes, and with the most accurate experimental evidence.

\end{abstract}

\pacs{47.27.ek, 64.60.Fr, 66.10.Cb, 64.60.Ht}

\maketitle

\section{INTRODUCTION}
Recently, there has been a renewed interest in the critical
behavior of simple and complex fluids, both with respect to
liquid-gas transitions and demixing transitions in binary fluids.
\cite{1,2,3,4,5,6,7,8,9,10} 
There has been remarkable progress both in the
fuller theoretical understanding of the asymptotic critical region
and the types of correction to scaling that occur in these systems,
\cite{1} and of the nature of the crossover towards classical,
van der Waals-like behavior further away from the critical point.
\cite{2} Also, the values of the critical exponents and other
universal properties (such as critical amplitude ratios) are now
known with high precision, from a variety of techniques
(renormalization group,\cite{3} Monte Carlo simulations,
\cite{4,5} and high-temperature series expansions \cite{6}).
Particularly relevant in the present context are advances in
the finite-size scaling analysis of computer simulations of fluids,
\cite{7,8,9,10} which allow one to study both universal and
non-universal critical properties of various off-lattice models of
fluids with an accuracy that is competitive with the work on Ising
lattice models.\cite{4,5}

With respect to critical dynamics in fluids, the situation is much
less satisfactory even though precise experimental data were presented
a long time ago \cite{11} and theoretical
analyses invoking either approximations, such as mode coupling theory
\cite{12,13,14,15,16,17} or low-order renormalization group
expansions in $\epsilon = 4-d$, where $d$ is the dimensionality,
\cite{18,19,20,21} do exist. One should note that dynamic
universality classes encompass fewer systems than static ones:
\cite{19,20} while uniaxial ferromagnets, binary alloys, liquid-gas criticality
and demixing in binary fluids all belong to the same
universality class as far as their static critical behavior is
concerned, these systems belong to more than one dynamic
universality class. Thus, there is a clear need for more
theoretical analyses of critical dynamics.

Particularly scarce are simulations of the critical dynamics of
fluids: of recent works, we are aware only of a 
nonequilibrium Molecular Dynamics (NEMD) calculation for
a two-dimensional fluid in which heat conduction
near the critical point was studied,\cite{22} and of a similar
investigation by Chen {\it et al.} \cite{nn23} of a three-dimensional
Lennard-Jones single component fluid. Beyond that
Jagannathan and Yethiraj (JY) \cite{23} used 
a three-dimensional Widom-Rowlinson model \cite{24} to study
the inter-diffusional critical dynamics in a binary-fluid.
However, the conclusions of this
latter work have been seriously challenged.\cite{nn23,25}

In the present paper, we move to fill this gap by
presenting a comprehensive simulation study of critical dynamics in
fluids by studying a symmetric binary fluid Lennard-Jones mixture. In
previous work \cite{26,27,28} we have shown that the coexistence
curve, concentration susceptibility, interfacial tension between
coexisting liquid phases, pair correlation functions and static
and dynamic structure factors for this model can be reliably
estimated {\it via} a combination of semi-grandcanonical Monte Carlo
methods (SGMC) \cite{26,27,28,29,30,31,32,33} and microcanonical Molecular
Dynamics (MD) methods.\cite{34,35,36,37} Transport coefficients
such as self-diffusion and interdiffusion coefficients,\cite{26}
shear viscosity \cite{26} and bulk viscosity \cite{27} were
estimated {\it away} from the critical region.
Here, however, we expressly address the critical behavior of the model
and compare with theoretical expectations. Moreover, our study
strongly supports the challenge \cite{25} to the earlier study by JY\cite{23}
of the somewhat similar but less realistic Widom-Rowlinson model.

In the balance of this article, 
Section II presents the model and briefly reviews our simulation
methods. Section III discusses the static critical properties that
are extracted from the ``raw data'' by a finite-size scaling
analysis.\cite{38,39,40,41} Sec. IV then presents our results
on the selfdiffusion coefficients and the shear viscosity:
we discuss them in the light of
the theoretical predictions.\cite{12,13,14,15,16,17,18,19,20,21} 
The interdiffusional coefficient, which vanishes fairly strongly,
requires more detailed, specifically, finite-size scaling 
considerations, etc. which are presented in
Sec. V. The article concludes in Sec. VI with a brief 
summary and discussion.

\section{MODEL AND SIMULATION TECHNIQUES}
Following Refs. $27-29$ we consider a binary fluid of
point particles with pairwise interactions 
in a cubical box of finite volume $V=L^3$ subject to periodic
boundary conditions.
Starting from a full Lennard-Jones potential
\begin{equation}\label{eq1}
\Phi_{{\rm LJ}}(r_{ij})=4\varepsilon_{\alpha \beta} [(\sigma_{\alpha
\beta}/r_{ij})^{12}-(\sigma_{\alpha \beta}/r_{ij})^6]\;,
\end{equation}
we construct a truncated potential that is strictly zero for $r_{ij} \geq
r_c$ as follows,\cite{34}
\begin{equation}\label{eq2}
u(r_{ij})=\Phi_{{\rm LJ}}(r_{ij})-\Phi_{{\rm LJ}}(r_c)- (r_{ij}-r_c)\frac{d\Phi_{{\rm LJ}}}{dr_{ij}}
|_{r_{ij}=r_c},~ \mbox{for}~ r_{ij} \leq r_c .
\end{equation}
This form ensures that both the potential and the force are continuous at $r=r_c$. In the
previous work,\cite{26,27,28} the last term on the right-hand-side of
Eq.~(\ref{eq2}) was not included so that the force at $r_{ij}=r_c$
exhibited a jump, while only the potential was continuous. This is not desirable
when considering dynamic behavior, because, in a microcanonical simulation,
this results in a drift of the total energy.

The parameters in Eqs.~(\ref{eq1}) and (\ref{eq2}) were chosen as
\begin{equation}\label{eq3}
\sigma_{{\rm AA}} = \sigma _{{\rm BB}} = \sigma_{{\rm AB}} = \sigma\;,
\end{equation}
and, hence, we adopt $\sigma$ as our unit of length. The
cutoff $r_c$ is chosen as $r_c=2.5 \sigma$. As
previously,\cite{26,27,28} we take the total particle number
$N=N_{{\rm A}} +N_{{\rm B}}$ of the binary A+B mixture and the volume such that
the density $\rho^*=\rho\sigma^3=N\sigma^3/V=1$. This choice is convenient, since
the system is then in its liquid (rather than its vapor) phase and
crystallization is not yet a problem at the
temperatures of interest.
Finally, the reduced temperature $T^*$ and
energy parameters $\varepsilon_{\alpha \beta}$ are chosen as
\cite{26,27,28} 
\begin{equation}\label{eq4}
T^*=k_{{\rm B}}T/\varepsilon~~~ \mbox{and}~~~\varepsilon_{{\rm AA}}=
\varepsilon_{{\rm BB}}=2\varepsilon_{{\rm AB}}=\varepsilon.
\end{equation}

The system is equilibrated as follows. First, a Monte Carlo (MC)
run is performed in the canonical ensemble ($N_{\rm A}$=$N_{\rm B},V,T$), 
starting out from
particles at random positions in the simulation box. The MC
moves used are random displacements of randomly chosen 
single particles (selecting the trial
value of each new cartesian coordinate in the range $[-
\sigma/20,+\sigma/20]$ about its old value) 
and applying the standard Metropolis acceptance criterion.
\cite{30,31} These initial runs were carried out for $10^4$
Monte Carlo steps (MCS) per particle, for systems with particle numbers from $N=400$
to $N=12800$. Then equilibration is continued, using the
semi-grandcanonical Monte Carlo (SGMC) method.
\cite{26,27,28,29,30,31,32,33} After $10$ displacement
steps per particle $N/10$ particles are randomly chosen in succession and
an attempted identity switch 
is made, ${\rm A} \rightarrow {\rm B}$ or ${\rm B} \rightarrow {\rm A}$,
where both the energy change, $\Delta E$, and the chemical potential
difference, $\Delta \mu$, between A and B particles enters the
Boltzmann factor. However, we restrict attention here to the
special case $\Delta \mu =0$, which, for the symmetric
mixture considered, means that for $T > T_c$ 
we simulate states with an
average concentration $\langle x_{\rm A}\rangle = \langle x_{\rm B} \rangle =
1/2~$ (with $x_{\alpha}=N_{\alpha}/N$), 
while for $T<T_c$ we simulate states along either the A-rich
or the B-rich branch of the coexistence curve.

In the semi-grandcanonical ensemble, the concentration $x_{\rm A}$ is a
fluctuating variable, and hence the probability distribution
$P(x_{\rm A})$ can be recorded: this is particularly useful in the context of a
finite-size scaling analysis.\cite{38,39,40,41} In addition, use
of the semi-grandcanoncial ensemble for the study of static critical
properties in a binary fluid mixture is preferable since
critical slowing down is somewhat less severe than in
the canonical ensemble.
Critical slowing down limits the accuracy that can be
attained, since the statistical error scales like $1/\sqrt{n}$,
where $n$ is the number of statistically independent
configurations.\cite{33} The statistically independent
configurations used in the computation of averages must be
separated from each other along the stochastic (MC) or
deterministic (MD) trajectory of the simulation by a time interval
which is of order of the longest relaxation time in
the system.\cite{33} In a finite system at the demixing critical
point of a fluid binary mixture, this slowest relaxation time
scales with box dimension $L$ as
\begin{equation}\label{eq5}
\tau_{max} \propto L^z,
\end{equation}
where $z$ is a universal dynamic critical exponent, which
depends on the dynamic universality class.\cite{18,19,20,21} For the
SGMC algorithm, the order parameter (concentration difference
between A and B) is not a conserved variable, while the average
density $\rho =N/V$ is conserved. As a consequence,
this model belongs to ``class C\hspace{1mm}'' in the Hohenberg-Halperin
classification,\cite{19} and hence the dynamic exponent is
roughly $z=2$.\cite{18,19} If we performed MC simulations
in the $N_{\rm A} N_{\rm B}VT$ ensemble, the order parameter would be a
conserved variable, in addition to the density (``class D''
\cite{19}), and then the dynamic critical exponent is expected to
be significantly larger, $z=4-\eta$, where $\eta\; (\simeq 0.03$)
\cite{3,4,5} describes the spatial decay of correlations at criticality.
If one performs MD runs in a microcanonical $N_{\rm A}N_{\rm B}VE$ ensemble (with
$N_{\rm A}=N_{\rm B}$ and with an energy chosen so that the system is
precisely at the critical point), the dynamic exponent is
predicted to be $z\simeq 3$.\cite{16,17,18} In fact,
Jagannathan and Yethiraj \cite{23} used Eq.~(\ref{eq5}) in order to
explore the critical dynamics of their Widom-Rowlinson model.\cite{24}
Comparing the values of $z$ for
the three algorithms discussed above, it is clear why the SGMC
algorithm has an advantage, as far as static critical properties
are concerned.

For the study of dynamic critical properties, multiple
independent initial configurations were prepared,\cite{yy42} from SGMC runs
with $5\times 10^5$ MCS
(but excluding states from the first
$10^5$ MCS). Then a further thermalization for $2\times 10^5$ MD steps
was carried out in the $NVT$ ensemble using the
Andersen thermostat.\cite{34,35,36,37} For all MD runs, we always
chose the masses of the particles equal to each other,
$m_{\rm A}=m_{\rm B}=m$, and applied the standard velocity Verlet algorithm
\cite{34,35,36,37} with a time step $\delta t^*=0.01/\sqrt{48}$ where
$t^*=t/t_0$ with scale factor
\begin{equation}\label{scft0}
t_0=(m \sigma ^2/\varepsilon)^{1/2}. 
\end{equation}
The
final production runs in the microcanonical ensemble (where the
Andersen thermostat is switched off) used from about $10^6$ MD
steps for temperatures $T^*=1.5$ and higher, but up to $2.8$$\times$$10^6$  MD
steps for $T$ close to $T_c$ (where $T_c^*=1.4230
\pm 0.0005$, see below).
To avoid confusion, we note that the
different value of $T_c$ in the previous work,\cite{26,27,28}
namely, $T_c^* = 1.638 \pm 0.005$, arose from the different choice of
potential (the last term on the right hand side of Eq.~(\ref{eq2}) being absent
in Refs.~$27-29$).

\section{STATIC CRITICAL PROPERTIES}
\begin{figure}[htb]
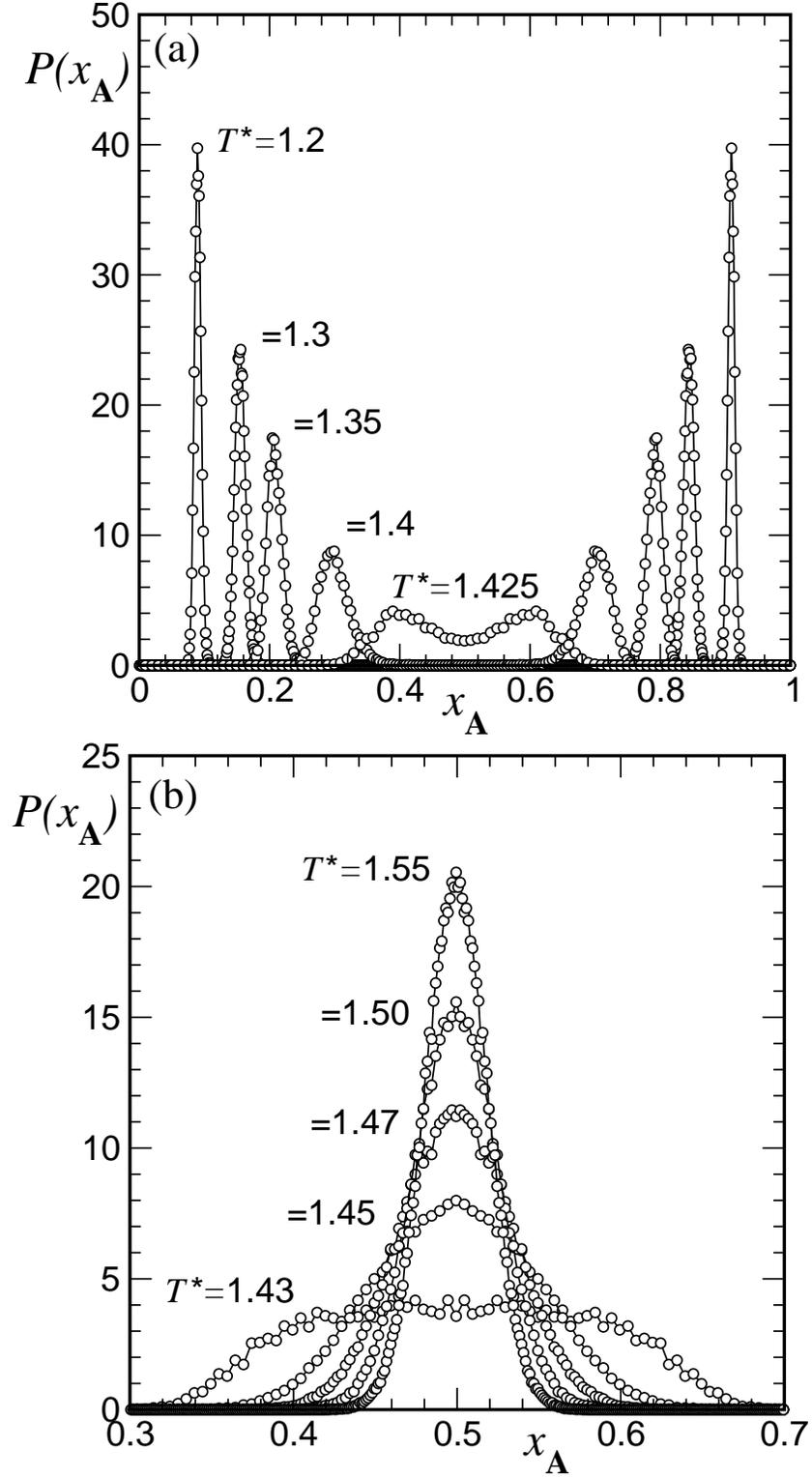

\centering
\includegraphics*[width=0.65\textwidth]{fig1a.eps}
\includegraphics*[width=0.67\textwidth]{fig1b.eps}
\caption{\label{fig1} Probability distributions $
P(x_{\rm A})$ of the relative concentration $x_{\rm A}=N_{\rm A}/N$ of A particles
for $N = 6400$ and chemical potential difference $\Delta \mu = 0$ at
several temperatures (a) below $T_c$ and (b) above $T_c$,
respectively. For clarity many independent data points have been omitted.}
\end{figure}

Using the SGMC algorithm we record the
fluctuating number of A particles $N_{\rm A}$ (recall that $N_{\rm B}=N-N_{\rm A}$
with the total particle number $N$ fixed) and generate histograms
to estimate the probability distribution $P(x_{\rm A})$ of the relative
concentration $x_{\rm A}=N_{\rm A}/N$. Typical ``raw data'' for $P(x_{\rm A})$ are
shown in Fig.~\ref{fig1}. The symmetry relation that holds for
$\Delta \mu =0$, namely,
\begin{equation}\label{eq6}
P(x_{\rm A})=P(1-x_{\rm A}),
\end{equation}
has been incorporated in the data. This has been done because 
below $T_c$, where $P(x_{\rm A})$
has a pronounced double-peak structure corresponding to the two
sides $x_{\rm A}^{{\rm coex}(1)},\; x_{\rm A}^{{\rm coex}(2)}$ of the two-phase
coexistence curve, transitions from one side to the other occur
very infrequently (or, at low temperatures, not at all). 
In Fig.~\ref{fig1}({\rm a}) we present probability distributions for
several temperatures below $T_c$ while Fig.~\ref{fig1}({\rm b}) shows the
distributions for temperatures above $T_c$. From
$P(x_{\rm A})$, we define the truncated moments $\langle x_{\rm A}^k \rangle$ of the
concentration distribution as follows \cite{26}
\begin{equation}\label{eq7}
\langle x_{\rm A}\rangle = 2 \int _{1/2} ^1 x_{\rm A}P(x_{\rm A})dx_{\rm A}\;,
\end{equation}
\begin{equation}\label{eq8}
\langle x_{\rm A}^k\rangle = 2 \int _{1/2} ^1 x_{\rm A}^kP(x_{\rm A})dx_{\rm A}\;.
\end{equation}
The two branches of the coexistence curve can then be estimated
for large $N$ via
\begin{equation}\label{eq9}
x_{\rm A}^{{\rm coex}(1)}\simeq (1-\langle x_{\rm A}\rangle),
\quad x_{\rm A}^{{\rm coex}(2)}\simeq \langle
x_{\rm A}\rangle\;,
\end{equation}
while the ``concentration susceptibility'' $\chi$ and its dimensionless form, 
$\chi^*$, can be estimated
from
\begin{equation}\label{eq10}
k_{\rm B}T\chi = T^*\chi^* = N(\langle x_{\rm A}^2 \rangle - 1/4) ,\quad T>T_c.
\end{equation}
Another useful quantity is the fourth-order cumulant $U_L$, defined
by \cite{4,39,43}
\begin{equation}\label{eq12}
U_L(T) = \langle (x_{\rm A}-1/2 )^4\rangle / [ \langle (x_{\rm A} -
1/2) ^2\rangle ^2].
\end{equation}

Note that for a finite system, $(\langle x_A \rangle-1/2)$ remains
nonzero even for $T \geq T_c$ [as seen in Fig.~\ref{fig1}(b)]. 
Furthermore $\langle x_A\rangle$ 
is a smooth function of temperature for finite $L$ while 
$\chi$ likewise remains finite at $T_c$. Due to these
effects, a finite-size scaling analysis of the SGMC
data for these quantities is clearly required, as is well known.
\cite{33,39,41} 

There are different strategies used in the literature to estimate
the location of the critical temperature $T_c$ from such
simulation results.\cite{4,5,7,8,9,10,33,36,39,41} The simplest
method, most often used for fluids, records
$x_{\rm A}^{{\rm coex}(1)},x^{{\rm coex}(2)}_{\rm A}$ for several choices of $L$ and
checks for a regime of temperatures below $T_c$ where the results
are independent of $L$ within statistical errors. In this regime
one chooses several temperatures, as close to $T_c$ as possible,
and fits to a power law
\begin{equation}\label{eq13}
x_{\rm A}^{{\rm coex}(2)}-x_{\rm A}^{{\rm coex}(1)}= B(1-T/T_c)^\beta\;,
\end{equation}
where the critical amplitude $B$ and $T_c$ are adjustable
parameters, while the critical exponent $\beta$ is fixed at its
theoretical value for the universality class of the
three-dimensional Ising model, $\beta \simeq 0.325$.
\cite{3,4,5,6} 

\begin{figure}[htb]
\centering
\includegraphics*[width=1\textwidth]{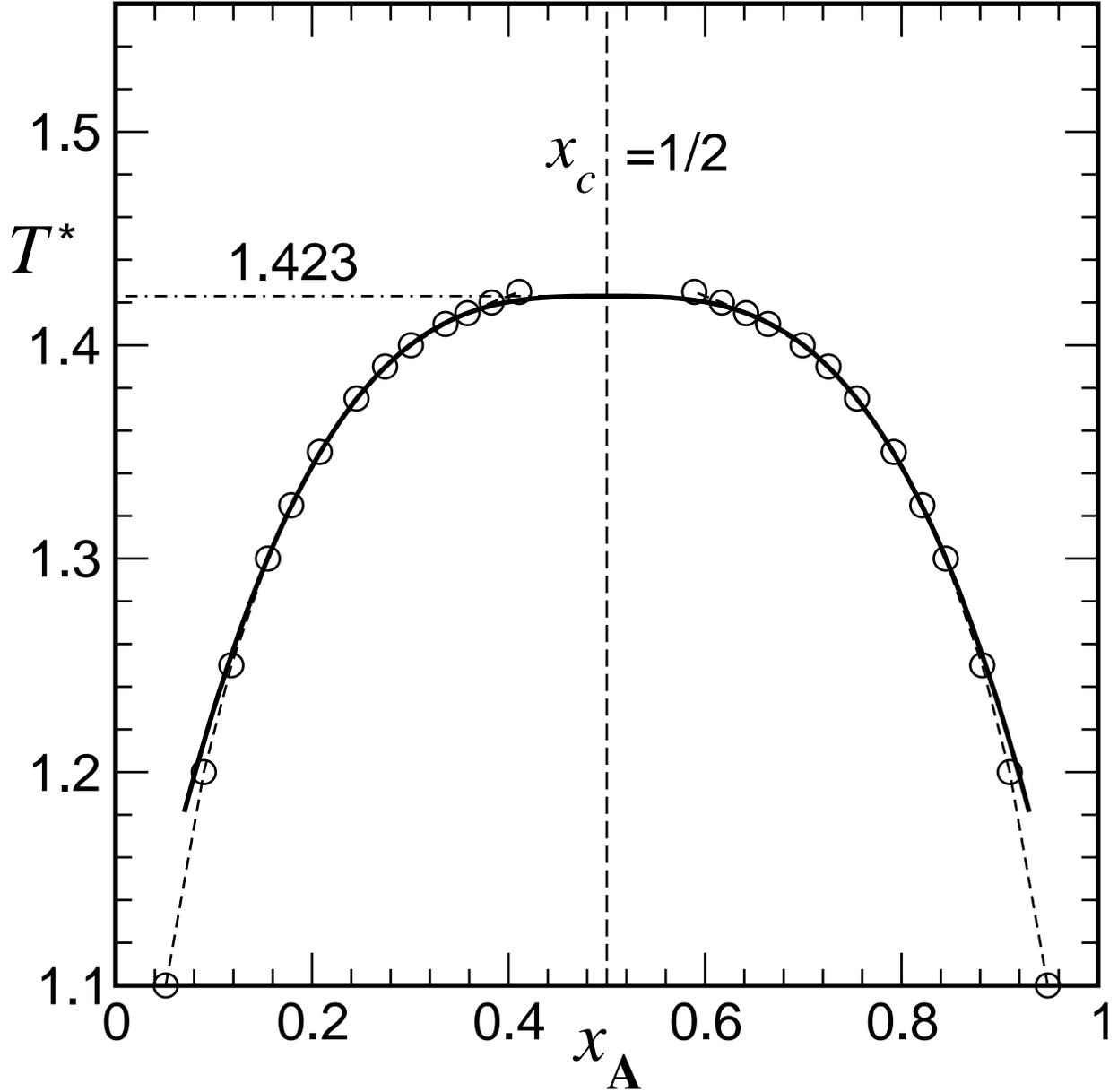}
\caption{\label{fig2}Coexistence curve of the symmetrical (truncated)
Lennard-Jones binary fluid in the plane of temperature $T$
and concentration $x_{\rm A}=N_{\rm A}/N$, for overall density $\rho^*=1.0$, 
the precise choice of potentials being given in
Eqs.~(\ref{eq1})-(\ref{eq4}). Open circles are the simulation results
for a system of $N=6400$ particles,
while the broken curve is only a guide to the eye. The solid curve
indicates a fit to Eq.~(\ref{eq13}) which yields $T_c^*=1.423$ as
highlighted by the horizontal dot-dashed line.}
\end{figure}

In Fig.~\ref{fig2}, we show the two-phase coexistence 
curve for $N=6400$. (Recall that for our choice we have density
$\rho^*=1$, so the system size is $L=N^{1/3}\sigma$.) The dashed line 
in Fig.~\ref{fig2} is a guide to the
eye for the numerical data (open circles). The continuous line is a fit to 
Eq. (\ref{eq13}) using the range from $0.2<x_{\rm A}<0.5$ (but excluding the two 
points closest to $T_c$). From this we obtain $T_c^*=1.423\pm0.002$ and
$B=1.53\pm 0.05$. This fit is good over a relatively
wide range of temperature but,
in reality, the range over which
Eq.~(\ref{eq13}) should be valid is significantly smaller owing to
various corrections to
scaling\cite{3,4,5,6} which have been neglected.
Hence, systematic uncontrollable errors
easily arise: the true value of $T_c$ could well be somewhat
lower with $B$ larger. Alternative methods are thus needed
to obtain more reliable estimates of $T_c$ with 
improved error bounds. Indeed one clear drawback of the previous study on critical
dynamics by JY\cite{23} is that the accuracy with which the critical
point could be located was relatively poor.

A method used more recently in Monte Carlo studies of
critical phenomena \cite{36,39,41} is data collapsing based on
the finite-size scaling hypothesis.\cite{38,41} Let us consider the concentration 
susceptibility $\chi$ in the vicinity of the critical point. At the
critical concentration, $x_A=x_B=1/2$, we have
\begin{equation}\label{singchi}
\chi^*(T) \approx \Gamma_0\epsilon^{-\gamma}~~~\mbox{with}~~~\epsilon=(T-T_c)/T_c,
\end{equation}
where $\Gamma_0$ and $\gamma$ are the critical amplitude and critical 
exponent, respectively.
From here on we shall use the symbol $\epsilon$ for the reduced temperature deviation
to avoid confusion with $t$ which arises naturally as a symbol for time
in the context of MD simulations.
For a system of finite size $L$, the basic
scaling ansatz may be written as
\begin{equation}\label{scachi}
\chi_L^* (T) \approx \Gamma_0Z(y)\epsilon^{-\gamma},
\end{equation}
where $y=L/\xi$, in which $\xi$ is the correlation length,
while $Z(y)$ is the appropriate finite-size scaling function. 
The correlation length diverges at criticality as 
\begin{equation}\label{singxi}
\xi\approx \xi_0\epsilon^{-\nu},
\end{equation}
with amplitude $\xi_0$ and exponent $\nu$.
In the limit
$y\rightarrow \infty$ (so that $L\rightarrow\infty$ at fixed $\epsilon > 0$)
the scaling function 
$Z(y)$ must approach unity so that Eq. (\ref{singchi}) is recovered.
For static quantities (such as $\chi$), in short-range
systems with periodic boundary conditions,
$Z(y)$ generally approaches unity exponentially fast. So, one may expect the
behavior
\begin{equation}\label{scfninfty}
Z(y)= 1+Z_{\infty}y^{\psi}e^{-ny}+..., \quad \mbox{for}\quad
y\rightarrow\infty,
\end{equation}
where the values of the exponent $\psi$ and the integer $n=1,2,3,...$ 
depend upon the details of the system
in question. On the other hand, for finite $L$, in the limit $y\rightarrow 0$
(so $\epsilon\rightarrow 0$ at fixed $L<\infty$), 
the susceptibility $\chi_L(T)$ is finite and its variation with $T$
must be smooth and analytic. Thus one should have 
\begin{equation}\label{scfnzero}
Z(y)= y^{\gamma/\nu}[Z_0+Z_1y^{1/\nu}+Z_2y^{2/\nu}+...~]\quad \mbox{as}\quad y\rightarrow 0.
\end{equation}

\begin{figure}[htb]
\centering
\includegraphics*[width=0.95\textwidth]{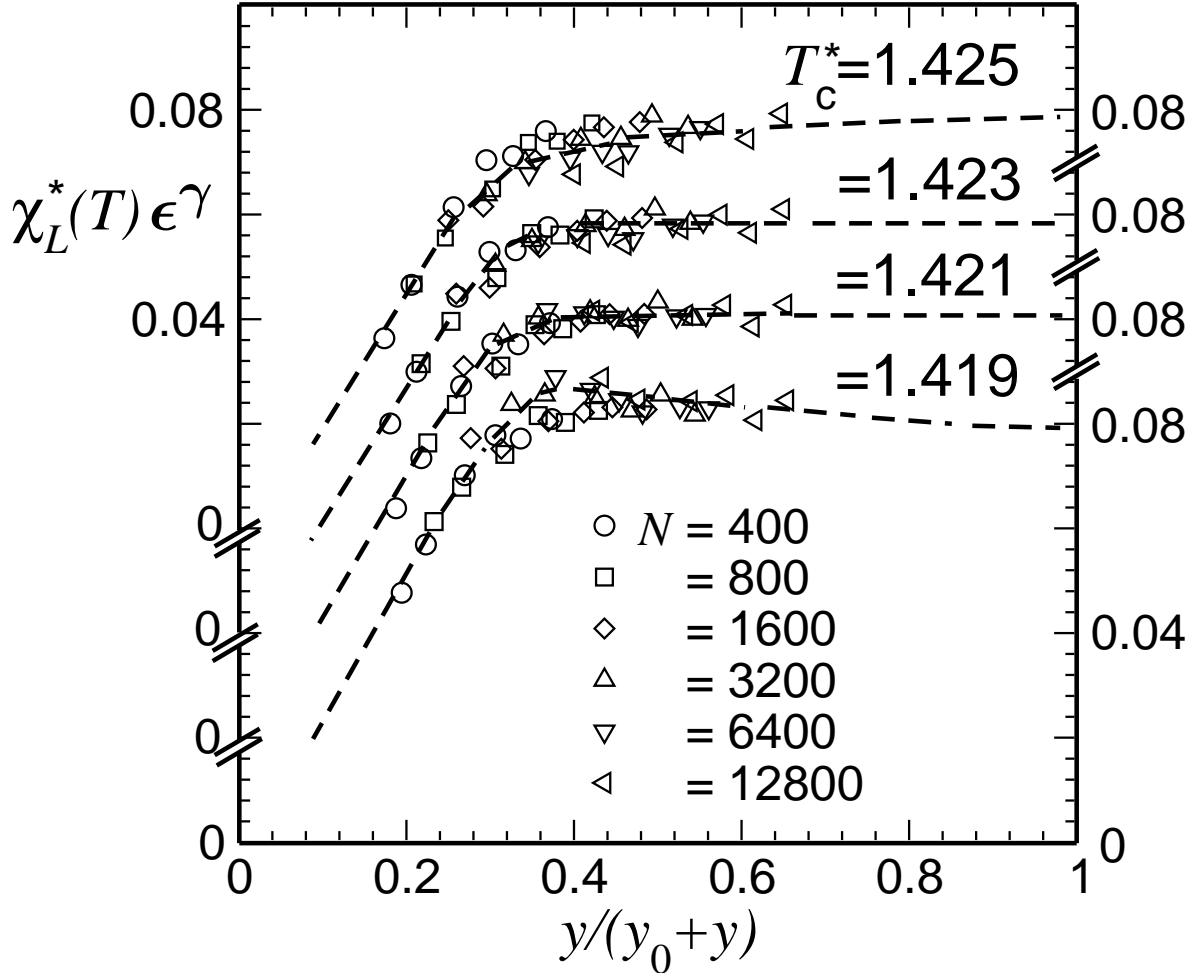}
\caption{\label{fig3}Finite-size scaling plots of the
susceptibility $\chi^*$ for temperatures above $T_c$ using the trial values of
$T_c^*$ marked in the figure.
The Ising values $\gamma=1.239$, $\nu=0.629$, have been accepted
and simulation results for
$\chi^*$ at temperatures $T^*=1.45,
1.46,1.48,1.50,1.52$, and $1.55$, are presented. Particle
numbers from $N=400$ to $N=12800$ are included, as indicated
(while the linear dimensions of the simulation box are $L=N^{1/3}\sigma$).
The dashed lines are guides to the eye: in light of the degree of data
collapse and the expected scaling function behavior stated in Eq.~(\ref{scfninfty}), the
estimates $T_c^*=1.423$ and $1.421$ are quite acceptable.}
\end{figure}

An effective procedure is then to study the computed quantity 
$\chi_L^*(T)\epsilon^{\gamma}$
as a function of $y$ for different system sizes by using $T_c$  as an 
adjustable parameter to
optimize the data collapse. 
The plot should then approach $\Gamma_0 Z(y)$.
From the asymptotic behavior of $\Gamma_0 Z(y)$ as $y$ becomes large
one can then estimate the critical amplitude. In Fig. \ref{fig3} we plot
$\chi_L^*(T)\epsilon^{\gamma}$ vs. $y/(y+y_0)$ noting that the abscissa variable approaches zero
when $y\rightarrow 0$ but tends to unity when $y\rightarrow\infty$ for $y_0>0$.
For convenience we have chosen $y_0=7$ which is comparable with the value of
$L/\xi$ for the largest system size. (This point will be discussed further.)
Of course, if one wishes to estimate all three quantities $T_c$,
$\gamma$ and $\nu$ from such a procedure, one again fights hard to 
control systematic errors since the uncertainities in the estimates for
these quantities are inevitably highly correlated.
Accordingly we have fixed the
critical exponents at their universal values, accurately known
from other studies,\cite{3,4,5,6} $\gamma \simeq 1.239, \nu
\simeq 0.629$, since there is no reason to doubt that these
exponents describe the static critical behavior of the present
model. 

In Fig~\ref{fig3}, we demonstrate data collapse for four 
trial values of $T_c$. It is clear that the collapse is 
inferior for the values $T_c^*=1.425$ and $1.419$
compared to the choices
$1.423$ and $1.421$: the collapse looks quite acceptable in these
latter cases. Thus, we conclude that our previous estimation 
of $T_c$ is consistent with this analysis.

An unbiased method to evaluate $T_c$ utilizes the fourth-order
cumulant defined in Eq.(~\ref{eq12}):\cite{39,43} plotting $U_L(T)$ vs. $T$
for various sizes $L$ one finds $T_c$ from a common intersection point
of these curves once corrections to finite-size scaling become
negligible. Quite generally one has
$U_L \rightarrow 1/3$ in any one-phase region while $U_L \rightarrow 1$
on the coexistence-curve diameter
in the two-phase region. 
Furthermore, for the three-dimensional Ising universality class 
$U_L(T_c)$ takes the value $0.6236$.\cite{8,9}

\begin{figure}[htb]
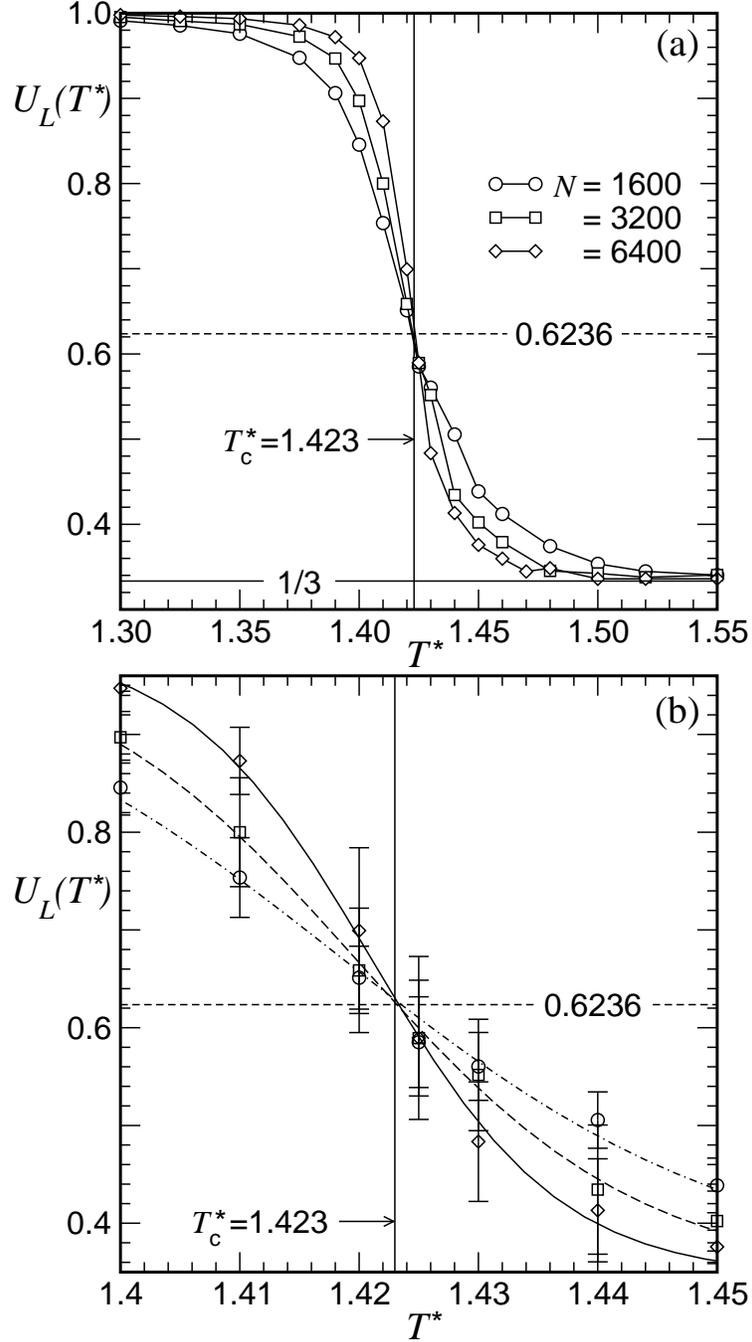

\centering
\includegraphics*[width=0.6\textwidth]{fig4a.eps}
\includegraphics*[width=0.6\textwidth]{fig4b.eps}
\caption{\label{fig4} The fourth-order cumulant $U_L(T)$ plotted vs.
$T$ for several system sizes, as indicated in the figure.
The broken horizontal line indicates the value of the
$U_L$ at $T_c$ for Ising type systems. 
The vertical line at $T^*=1.423$ represents our preferred estimate of $T_c^*$.
The smooth curves in the enlarged plot (b) are fits to tanh functions.}
\end{figure}

In Fig.~\ref{fig4}(a) we present $U_L(T)$
for several system sizes (as indicated in the figure) over a
rather wide range of temperature. The horizontal dashed line indicates
the value of this quantity at the critical point for the Ising
universality class. This plot clearly confirms that our model
belongs to the three-dimensional Ising universality class. 
While on a coarse scale the expected intersection is nicely seen,
the enlarged view of the data, in Fig~\ref{fig4}(b), 
reveals some scatter, which is mainly
due to the statistical
errors of the simulation data 
[as can be seen, by comparing with plots in Refs. $8,9$(c) and $45$]. 
In light of these statistical errors, further
analyses are clearly not warranted. Thus, for example, the
method of Wilding \cite{7} based on the use of the full
distribution $P(x_A)$ at criticality, rather than using only the
second and fourth moment, is not tried here:
but see also the critique in Ref. $46$. In Fig.~\ref{fig4}(b),
the smooth lines are fits to the hyperbolic tangent function. For the
system sizes shown, these fits all intersect one another close to
$T^*= 1.423$ and at the Ising critical value \cite{4} $0.6236$. Since this method appears to be the
most reliable currently available in the literature, we adopt
\begin{equation}\label{finaltc}
T_c^*=1.4230\pm0.0005
\end{equation}
for the subsequent analysis of our simulation data.

\begin{figure}[htb]
\centering
\includegraphics*[width=0.9\textwidth]{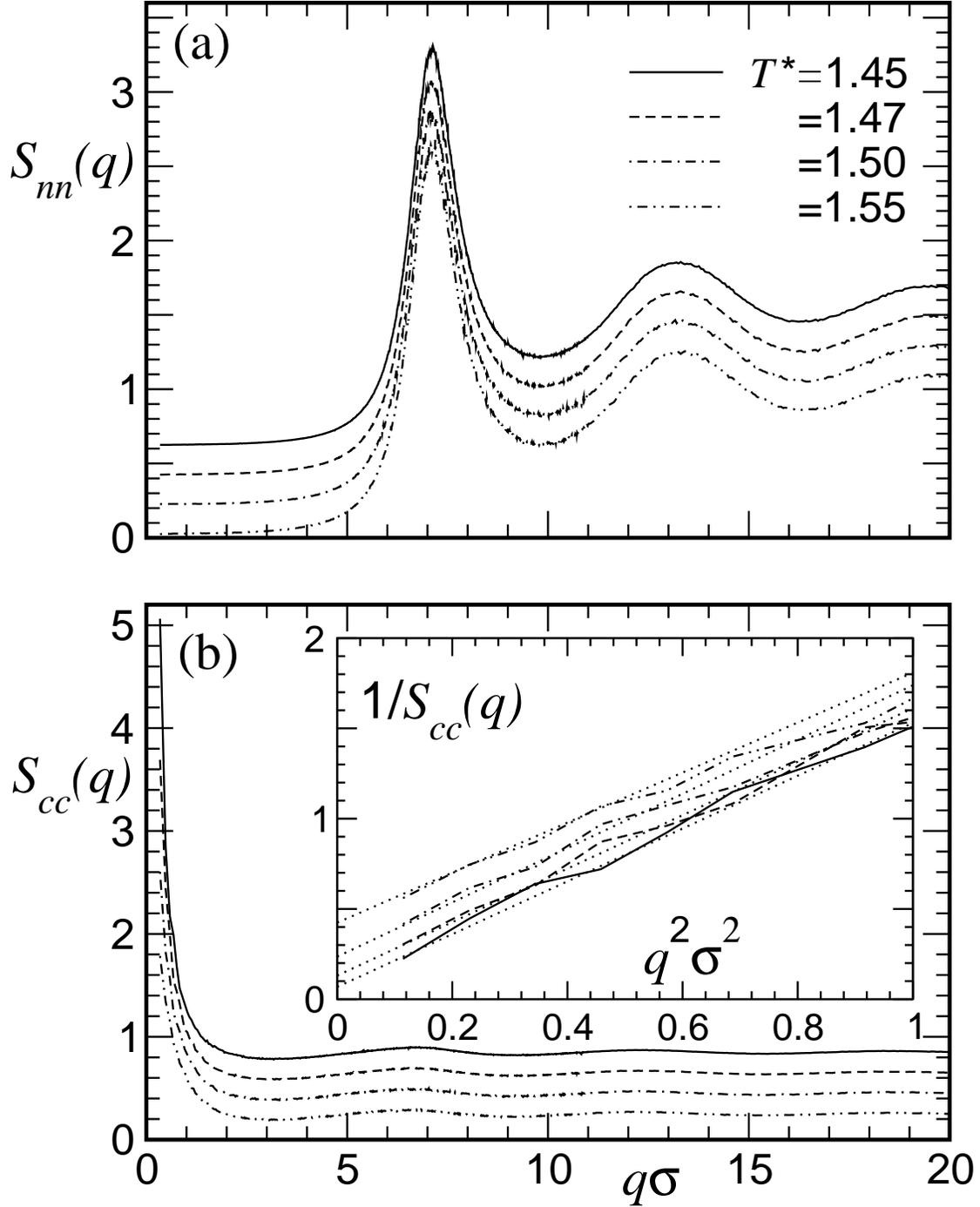}
\caption{\label{fig5} Plot of the structure
factors (a) $S_{nn}(q)$, (b) $S_{cc}(q)$, for various temperatures,
versus momentum $q$. The various curves are shifted up by 0.2 relative to
one another for clarity. 
All data refer to a system of $N=6400$ particles. Inset in part (b) 
represents an Ornstein-Zernike plot which yields estimates for $\xi(T)$
via Eq.~(\ref{eq18}).}
\end{figure}

Of course, it is also of interest to extract the pair correlation
functions $g_{{\rm AA}}(r),\; g_{{\rm AB}}(r)$ and $g_{{\rm BB}}(r)$ and their
Fourier transforms, $S_{{\rm AA}}(q),\; S_{{\rm AB}}(q)$ and $S_{{\rm BB}}(q)$, as
described for a closely related model (outside the critical
region) in earlier work.\cite{26} Of particular interest are
the combinations that single out number-density fluctuations
$S_{nn}(q)$ and concentration fluctuations $S_{cc}(q)$, defined via
\cite{44}
\begin{equation}\label{eq16}
S_{nn}(q) = S_{{\rm AA}}(q) +2S_{{\rm AB}}(q) + S_{{\rm BB}} (q)\;,
\end{equation}
\begin{equation}\label{eq1^7}
S_{cc}(q)=(1-x_{\rm A})^2S_{{\rm AA}}(q)+x^2_{\rm A}S_{{\rm BB}}(q)-2x_{\rm A}(1-x_{\rm A})S_{{\rm AB}}(q)\;.
\end{equation}
Fig.~\ref{fig5}(a) shows that $S_{nn}(q)$ exhibits the normal oscillatory behavior of
the structure factor of a dense liquid: the approach to
criticality has little discernible effect. By contrast, $S_{cc}(q)$ varies weakly
at large $q$, corresponding to small spatial length scales,
while at small $q$ a strong increase is observed. 
Of course, this is the expected Ornstein-Zernike behavior reflecting the
``critical opalescence'' due to concentration
fluctuations: see Fig.~\ref{fig5}(b). The inset here  displays an
Ornstein-Zernike plot, based on
\begin{equation}\label{eq18}
S_{cc}(q)=k_{\rm B}T\chi/[1+q^2\xi^2+...],
\end{equation}
from which our estimates of
the correlation length $\xi$ have been extracted. Note that, in the fitting
process, we have used the value of $k_{\rm B}T\chi$ from our SGMC simulations. 

\begin{figure}[htb]
\centering
\includegraphics*[width=0.7\textwidth]{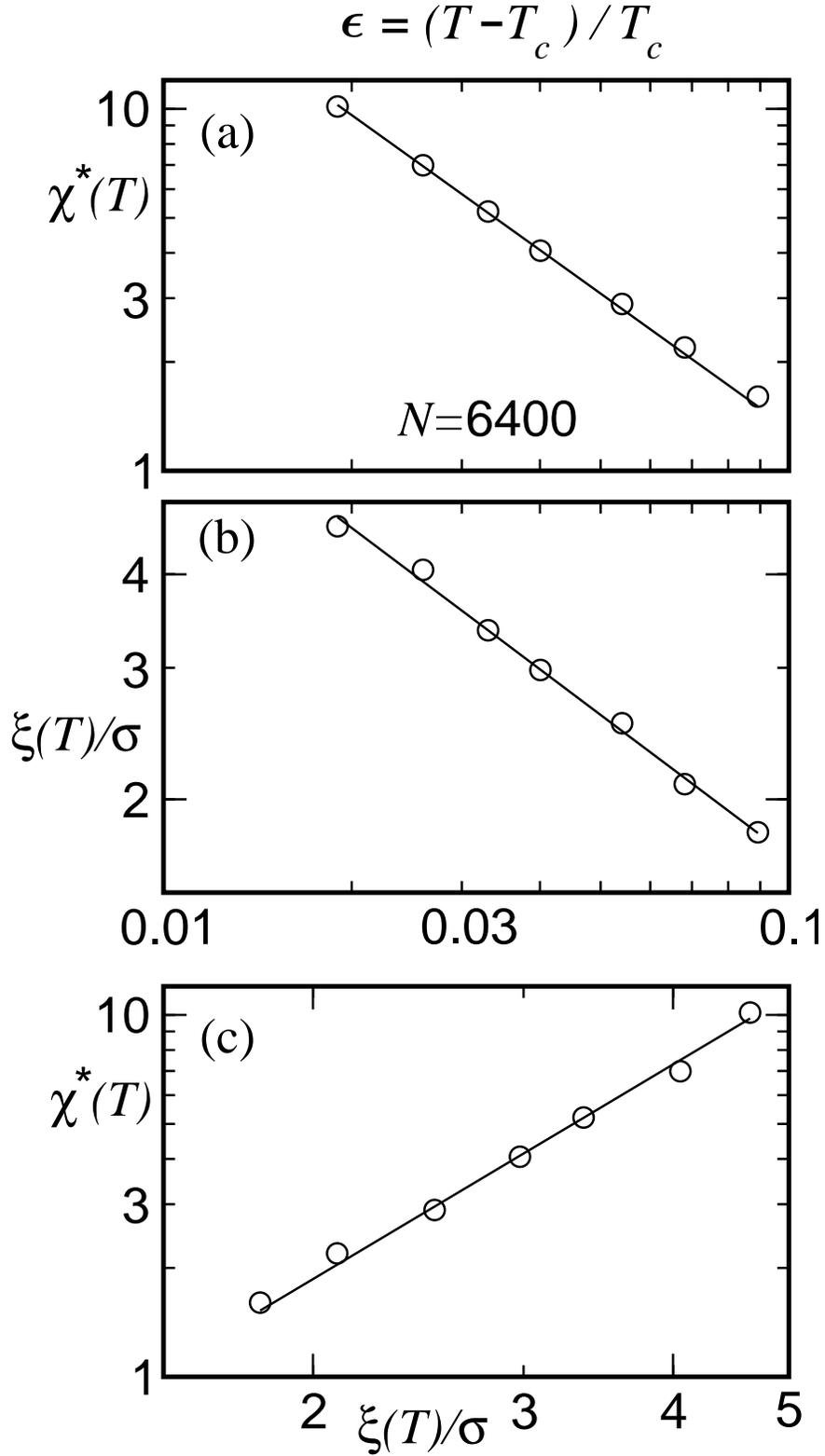}
\caption{\label{fig6} Plots of (a) the reduced
susceptibility $\chi^*$ and (b) the correlation length $\xi$ 
versus $\epsilon$. Part (c) shows the variation of $\chi$ with
$\xi$. The lines represent fits using the anticipated Ising exponents.
All the data refer to systems of $N=6400$ particles.}
\end{figure}

In Fig.~\ref{fig6} we plot the susceptibility and correlation length $\chi^*(T)$  and $\xi(T)$,  
as functions of $\epsilon$ for $N=6400$ and fit the data with the respective 
asymptotic forms (\ref{singchi}) and (\ref{singxi}).
For the fitting we again adopt the Ising critical
exponent values, so that 
the amplitudes are the
only adjustable parameters. The quality of the fits suggests that
the finite-size effects are negligible in this temperature range
(where as one sees from Fig.~\ref{fig3}, $y\gtrsim 4$ so that $L\gtrsim 4\xi(T)$).
The amplitudes are found to be
\begin{equation}\label{xi0}
\xi_0/\sigma=0.395\pm 0.025,~~ \Gamma_0=0.076\pm 0.006.
\end{equation}
In Fig.~\ref{fig6}(c)
we plot $\chi$ vs. $\xi$. The continuous line is a power-law fit with
the exponent $\gamma/\nu\simeq 1.970$.

We conclude this section by noting that no unexpected features have been uncovered.
The static properties comply fully with the anticipated critical
behavior characterizing three-dimensional Ising-type 
systems. Furthermore, the corrections to
scaling seem to be quite small in the temperature range covered by our simulations, 
so that one can observe the asymptotic
power laws even relatively far from the critical temperature.

\section{SELF-DIFFUSION COEFFICIENT AND SHEAR VISCOSITY NEAR CRITICALITY}
\begin{figure}[htb]
\centering
\includegraphics*[width=0.59\textwidth]{fig7a.eps}
\includegraphics*[width=0.625\textwidth]{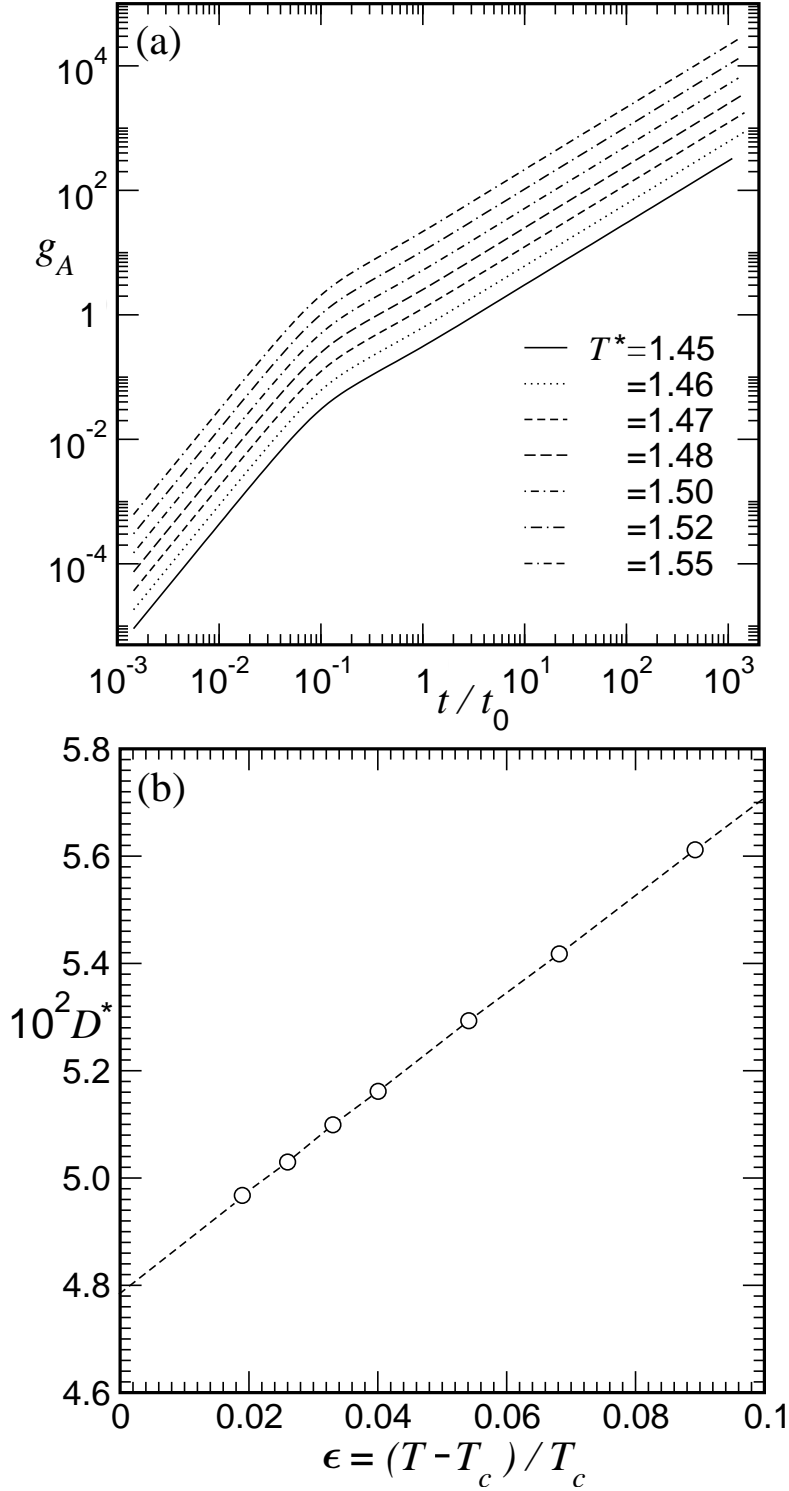}
\caption{\label{fig7}(a) Log-log plot of the mean
square displacements of all the particles versus time with
$t_0=(m\sigma ^2/
\varepsilon)^{1/2}$, for systems containing $N=6400$
particles, at the critical concentration and the seven temperatures
indicated. The plots for different $T$ are displaced by factors of $2$.
(b) Variation of the reduced self-diffusion constant $D^*$ with temperature.}
\end{figure}

The transport coefficients which are most readily found from
simulations are the self-diffusion coefficients $D_{\rm A}$ and $D_{\rm B}$ which can
be extracted from the Einstein relations for the mean square
displacements of tagged particles, \cite{xx45}
namely,
\begin{equation}\label{eq19}
g_{\rm A}(t)=\langle
[\vec{r}_{i,{\rm A}}(0)-\vec{r}_{i,{\rm A}}(t)]^2\rangle,
\end{equation}
and likewise for $g_{\rm B}(t)$,
where it is understood that the average $\langle \ldots \rangle$
includes an average over all particles of type A or B,
respectively. The self-diffusion coefficients then follow
from
\begin{equation}\label{eq20}
D_{\rm A}^* = (t_0/\sigma^2)D_{\rm A} = (t_0/\sigma^2) \lim_{t \rightarrow \infty} [g_{\rm A}(t)/6t],
\end{equation}
and similarly for $D_{\rm B}$. 
In the region above $T_c$
and at critical concentration $x_{\rm A}=x_{\rm B}=1/2$, to which we will restrict attention,
the symmetry of the
model requires $g_{\rm A}(t)=g_{\rm B}(t)$ and $D_{\rm A}=D_{\rm B}=D$. 
To extract $D$ and study its temperature dependence we have
hence averaged over the mean square displacements of all particles:
see Fig.~\ref{fig7}. Note that in the initial, ballistic regime $g_{\rm A}(t)$
varies quadratically with $t$ before crossing over to linear diffusive
behavior from which $D_{\rm A}$ is estimated.
As expected, the temperature
dependence of $D$ is rather weak and, indeed, close to linear
over this fairly narrow temperature range; moreover, $D$ appears to remain
nonzero at $T=T_c$ with a value close to $D^*=0.048$. Indeed, there
is no sign of any critical anomaly,
consistent with the previous work 
of JY.\cite{23} Similarly, a study of self-diffusion near the vapor-liquid
critical point of a lattice gas model \cite{45} did not detect any
significant critical anomaly. (Note, however, that this latter  model 
belongs to class B in the Hohenberg-Halperin classification.
\cite{19}) Nevertheless, a weak anomalous decrease of the self-diffusion coefficient at the
critical density has been seen in simulations of simple fluids
near the vapor-liquid critical point,\cite{46} but has not yet been confirmed
experimentally.\cite{xx47}

Next we consider the reduced shear viscosity $\eta^*$ which we calculate
from the Green-Kubo formula \cite{47}
\begin{equation}
\eta^*(T)=(t_0^3/\sigma Vm^2T^*)\int_{0}^{\infty} dt\langle\sigma_{xy}(0)
\sigma_{xy}(t)\rangle,
\end{equation}
where the pressure tensor $\sigma_{xy}(t)$ is given by
\begin{equation}
\sigma_{xy}(t)=\sum_{i=1}^{N}[m_i v_{ix} v_{iy}+
\frac{1}{2}\sum_{j(\ne i)}|x_i-x_j|F_y(|\vec r_i-\vec r_j|)],
\end{equation}
in which $\vec v_i$ is the velocity of particle $i$ while
$\vec F$ is the force acting between particles
$i,j$. 

Theory \cite{14,15,16} suggests that at the critical point $\eta^*$ should diverge as
\begin{equation}\label{singeta}
\eta^* \approx \eta_0\epsilon^{-\nu x_{\eta}} \sim \xi^{x_{\eta}},
\end{equation}
where $\eta_0$ and $x_{\eta}>0$ are the appropriate critical amplitude and
exponent. Renormalization-group theory \cite{18,19}
predicts $x_{\eta}\simeq 0.065$ while the theory of Ferrell and Bhattacharjee \cite{48}
yields $x_{\eta}\simeq 0.068$. These values are consistent with experiments \cite{49,50} which
yield $x_{\eta}$ between $0.064$ and $0.070$. The most recent theoretical estimate is \cite{48}
$0.0679\pm0.0007$ and the most recent experimental value is \cite{49} $0.0690\pm0.0006$.

\begin{figure}[htb]
\centering
\includegraphics*[width=1\textwidth]{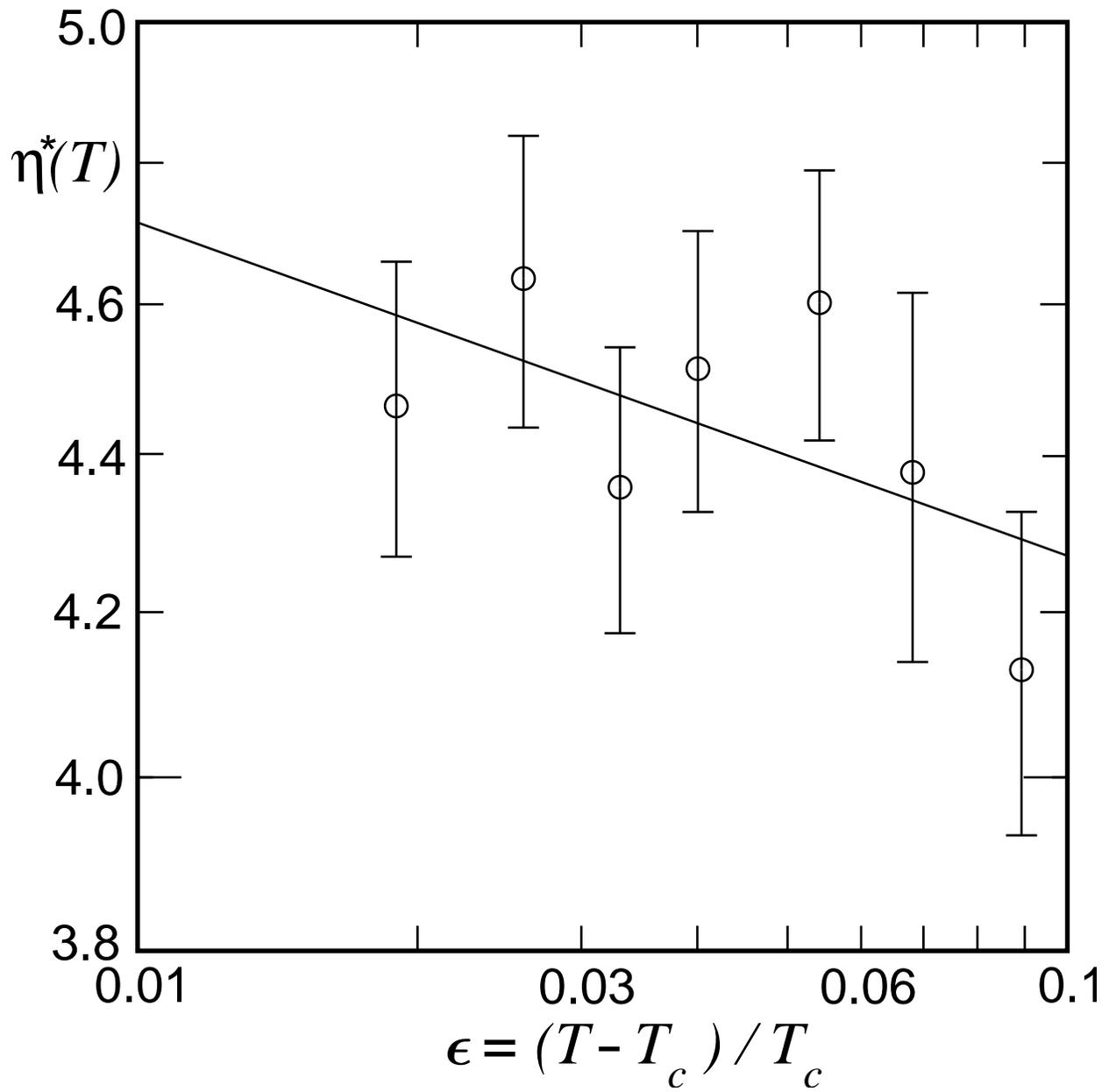}
\caption{\label{fig8} A log-log plot of the reduced shear viscosity $\eta^*$ vs. temperature.
The line represents a least squares fit to the theoretical form (\ref{singeta}) with 
$x_{\eta}=0.068$ and $\nu=0.629$, yielding an amplitude $\eta_0=3.8_7\pm0.3$.}
\end{figure}

Fig.~\ref{fig8} displays a log-log plot of $\eta^*$ vs. $\epsilon$. 
As normal in MD simulations, accurate estimation of the shear viscosity is difficult
and the large error bars shown in Fig.~\ref{fig8} prevent us from making
definitive statements about the critical singularity. But the
slow increase of $\eta^*$ as $T\rightarrow T_c$ 
is, in fact, compatible with the expected power law divergence, as Fig.~\ref{fig8}
shows, since the fitted line has a slope corresponding to $\nu=0.629$ and $x_{\eta}=0.068$
in Eq.~(\ref{singeta}). Although this fit
is consistent with the theoretical prediction, estimating $\nu x_{\eta}$ from the
data itself is clearly of little value. However, the amplitude, for which we obtain
$\eta_0=3.8_7\pm0.3$, will prove to be significant. 

\begin{figure}[htb]
\centering
\includegraphics*[width=1\textwidth]{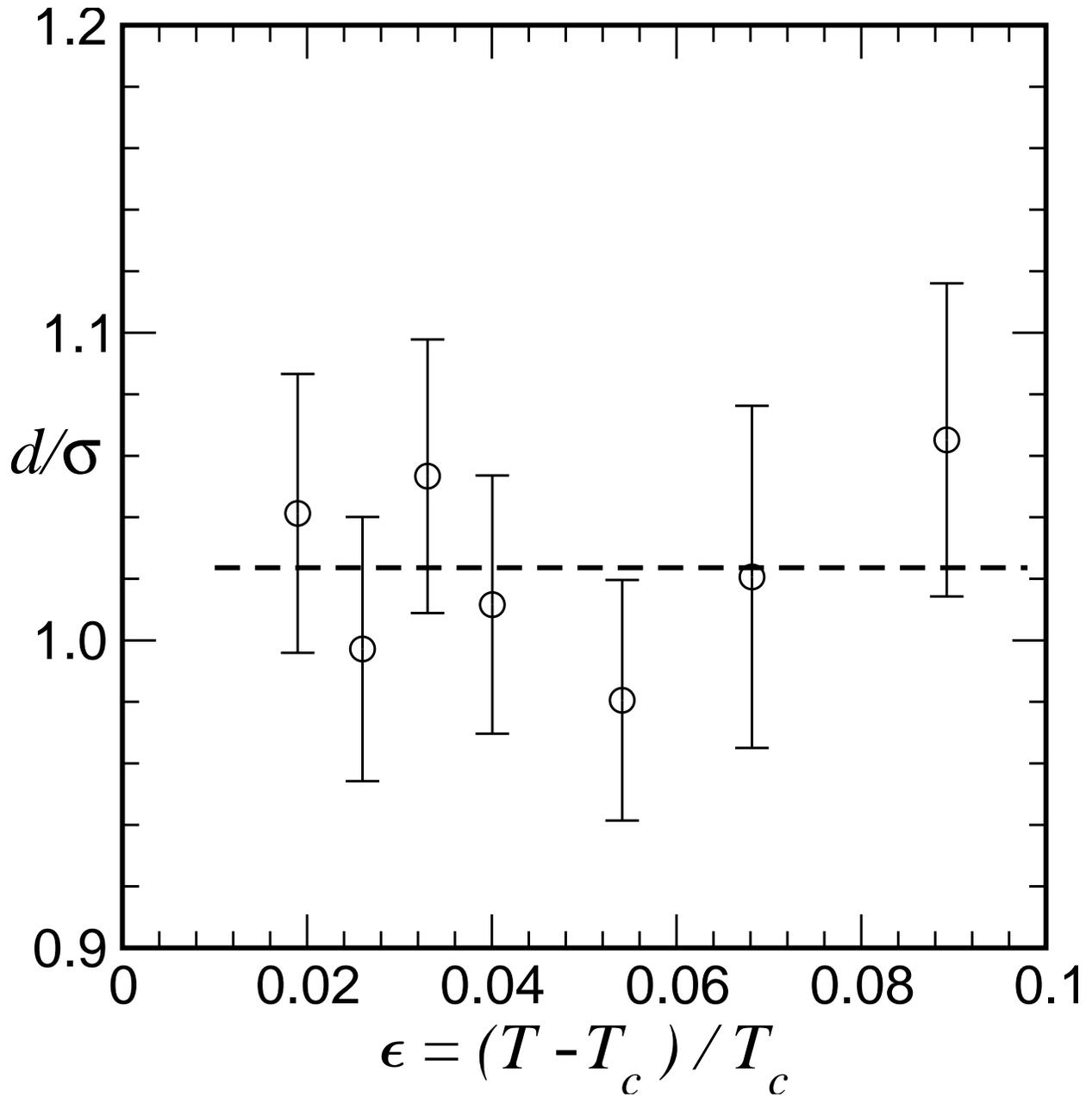}
\caption{\label{fig9} Plot of the Stokes-Einstein diameter, $d$, as defined in 
Eq.~(\ref{steselfd}), vs.
temperature. The dashed line serves as a guide to the eye.}
\end{figure}

At this point it is of interest to check the validity of the Stokes-Einstein relation,
which relates the self-diffusion constant of a diffusing spherical particle
of diameter $d$ to the shear viscosity of the fluid $\eta(T)$. For a particle moving in a fluid
of like particles, the Stokes-Einstein diameter $d$ can be written \cite{26}
\begin{equation}\label{steselfd}
d=\sigma T^*/2\pi\eta^* D^*,
\end{equation}
which corresponds to the assumption of slip boundary conditions on the surface of the
diffusing sphere. For stick boundary conditions, a factor $3$ replaces the factor $2$ in Eq.~(\ref{steselfd}).
In Fig.~\ref{fig9}, we present a plot of $d$ in the interval $T^*=1.45$ to $1.55$.
The data suggest that Eq.~(\ref{steselfd}) is still a valid approximation despite the strong
concentration fluctuations close to $T_c$. However, we do not expect the relation to remain valid closer
still to $T_c$ since $\eta(T)$ diverges, albeit slowly, while $D$ remains finite.

\section{INTERDIFFUSION:~~FINITE-SIZE SCALING}

Finally we consider the interdiffusion or mutual diffusion
coefficient, which is expected to vanish at the
critical point. Following the previous work \cite{26,27,28} we
use the Green-Kubo formula \cite{47} which we express as
\begin{equation}\label{eq21}
D_{{\rm AB}}(T)= \frac{\sigma^2}{t_0}D_{{\rm AB}}^*(T)=
\frac{\sigma^2}{t_0}\lim _{t \rightarrow \infty}D_{{\rm AB}}^*(T;t),
\end{equation}
where, introducing the appropriate reduced Onsager coefficient,
\begin{equation}
\mathcal L(T)=\lim _{t \rightarrow \infty} \mathcal L(T;t),
\end{equation}
we have the relation
\begin{equation}\label{dab2}
D_{{\rm AB}}^*(T;t)=\mathcal L(T;t)/\chi^*(T),
\end{equation}
in which, for numerical purposes, we will use our fits to Eq.~(\ref{singchi}) for $\chi^*$,
while
\begin{equation}\label{dab3}
\mathcal L(T;t)=(t_0/N\sigma^2T^*)\int_{0}^{t} dt'\langle J_x^{{\rm AB}}(0)J_x^{{\rm AB}}(t')\rangle,
\end{equation}
in which the current vector $\vec{J}^{{\rm AB}}$ is defined by
\begin{equation}\label{eq22}
\vec{J}^{{\rm AB}}(t)=x_{\rm B} \sum \limits _{i=1}^{N_{\rm A}}
\vec{v}_{i,{\rm A}}(t)-x_{\rm A} \sum\limits _{i=1}^{N_{\rm B}} \vec{v}_{i,{\rm B}}(t)\;,
\end{equation}
where $\vec{v}_{i,\alpha}(t)$ denotes the velocity of particle $i$
of type $\alpha$ at time $t$.

\begin{figure}[htb]
\centering
\includegraphics*[width=1\textwidth]{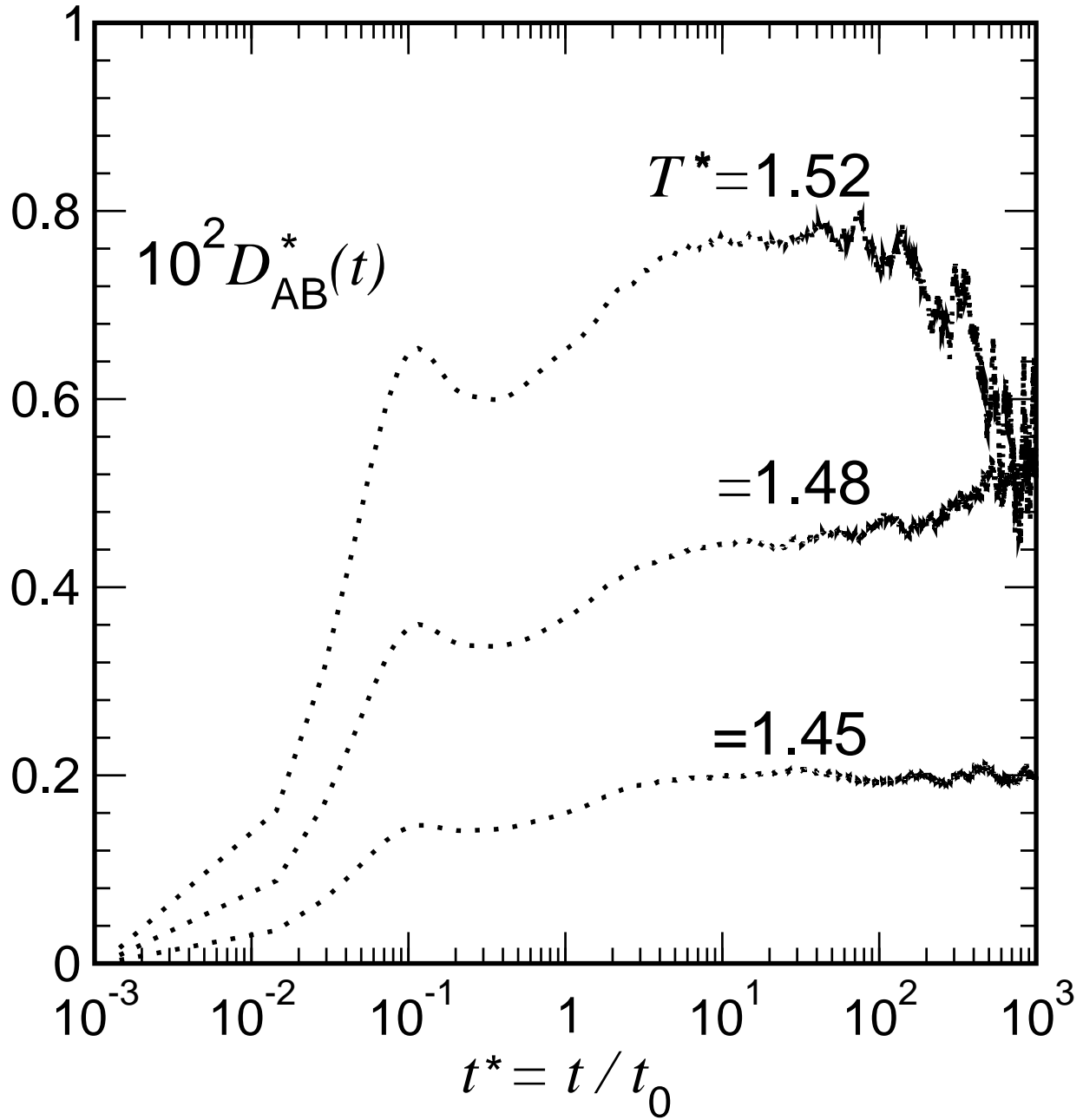}
\caption{\label{fig10} Plot of the interdiffusion coefficient
$D_{{\rm AB}}^*(t)$ vs. time at three different temperatures for
systems of $N=6400$ particles. The knees visible at short time are due to
discrete integration time step $\Delta t^*$.}
\end{figure}

Since the correlation function $\langle
J_x^{{\rm AB}}(0)J_x^{{\rm AB}}(t)\rangle$ is rather noisy at large times
comparable to the total observation time, $t_{{\rm obs}}$, of
an MD run, it is difficult to attain a precision
for $D_{{\rm AB}}$ as high as for the self-diffusion constant $D$.
This problem is evident in Fig.~\ref{fig10}, where short-time maxima
(seen for $t^* \simeq 0.1$) are followed by shallow minima (at $t^* \simeq 0.5-0.6$) that appear
before the expected plateau starts at about $t^*\simeq 4$.
For times $t^* \geq 40$, the curves 
become progressively more noisy, and, clearly, at most temperatures
the data for $t^* \geq 100$ may be discarded owing to deficient
statistics. But, in any case, the facts that the general shape of the
curves is similar for all temperatures studied, and, in
particular, that the time needed for $D_{{\rm AB}}(t)$ to reach a
plateau value is almost independent of $T$, suggest that
the dominant contribution to the temperature dependence of $D_{{\rm AB}}$ 
arises from $\chi^*$ in Eq.~(\ref{dab2}).

\begin{figure}[htb]
\centering
\includegraphics*[width=1\textwidth]{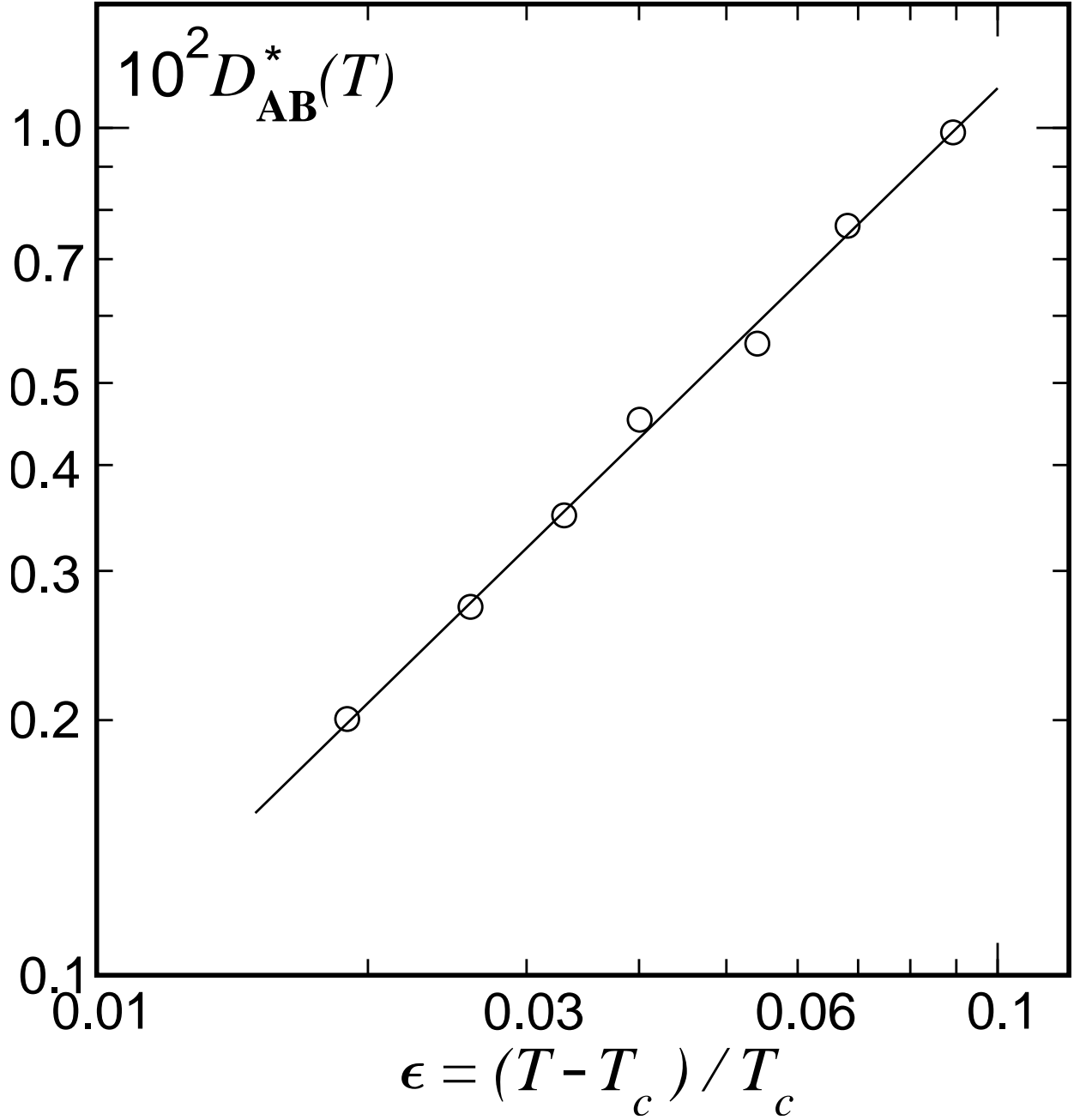}
\caption{\label{fig11} Log-log plot of the interdiffusion coefficient $D_{{\rm AB}}^*$ 
as calculated vs. $T$. The line is a fit to
the power law $D_{{\rm AB}}\sim \epsilon^{x_{{\rm eff}}\nu}$ which yields $x_{{\rm eff}}\simeq 1.6$.
The data correspond to
$N=6400$.}
\end{figure}

Fig.~\ref{fig11} presents plots of $D_{{\rm AB}}^*(T)$ versus
$T$ for systems of $N=6400$ particles. One sees
that $D_{{\rm AB}}$ has the apparent power-law behavior 
\begin{equation}\label{eq23}
D_{{\rm AB}} \sim \xi^{-x_{{\rm eff}}}\sim \epsilon^{x_{{\rm eff}}\nu}
\simeq \epsilon^1,~~ \mbox{with}~~ x_{{\rm eff}} \nu \simeq 1,~~ \mbox{so}~~ x_{{\rm eff}}\simeq
1.6.
\end{equation}
This result is in strong disagreement with the theoretical prediction for the 
interdiffusional critical exponent,\cite{48} namely,
\begin{equation}\label{xdth}
x_D=1+x_{\eta}\simeq 1.0679,
\end{equation}
that should be accessible asymptotically when $T\rightarrow T_c$.
Indeed, the value of $x_{{\rm eff}}$ found from Fig.~\ref{fig11} is
even larger than the value $x_D=1.26\pm0.08$ quoted by JY\cite{23}
for their binary fluid model. However, it would be quite erroneous to conclude
from our data and Fig.~\ref{fig11} that the simulations indicate a
serious failure of the theory. Even though we have found that the
finite-size effects in the equilibrium {\it static} properties are small 
for the temperature range and system sizes explored
[where $L\gtrsim 4\xi(T)$], one must be prepared to 
encounter much stronger finite-size corrections in {\it transport} properties
near criticality.

In addition experiments have shown that the Onsager coefficient near a
liquid-liquid demixing critical point or its equivalent, the thermal conductivity
near a vapor-liquid critical point, may have a significant noncritical
background contribution arising from short-range fluctuations.\cite{xx50,xx51}

Thus it is crucial to analyze dynamical simulations by making proper
allowance for the finite-size behavior and also
for possibly significant `background' contributions to the
quantities computed.\cite{new57} To this end we consider, as above, only the critical
isopleth $x_{\rm A}=x_{\rm B}=1/2$ and will focus on the finite-size Onsager coefficient
$\mathcal L_L(T)$ as defined via Eq.~(\ref{dab3}). The prime reason for
analyzing the Onsager coefficient rather than the interdiffusional
constant $D_{{\rm AB}}(T)$ (which clearly follows by dividing by $\chi$) is that
it represents most directly the basic fluctuation sum/integral analogous
to expressions like Eq.~(\ref{eq10}) defining $\chi$, or the standard
fluctuation sums for the specific heat, etc.; \cite{47} experience
teaches that such properties display the simplest, albeit not ``simple'',
singularity structure and finite-size behavior. 

\begin{figure}[htb]
\centering
\includegraphics*[width=1\textwidth]{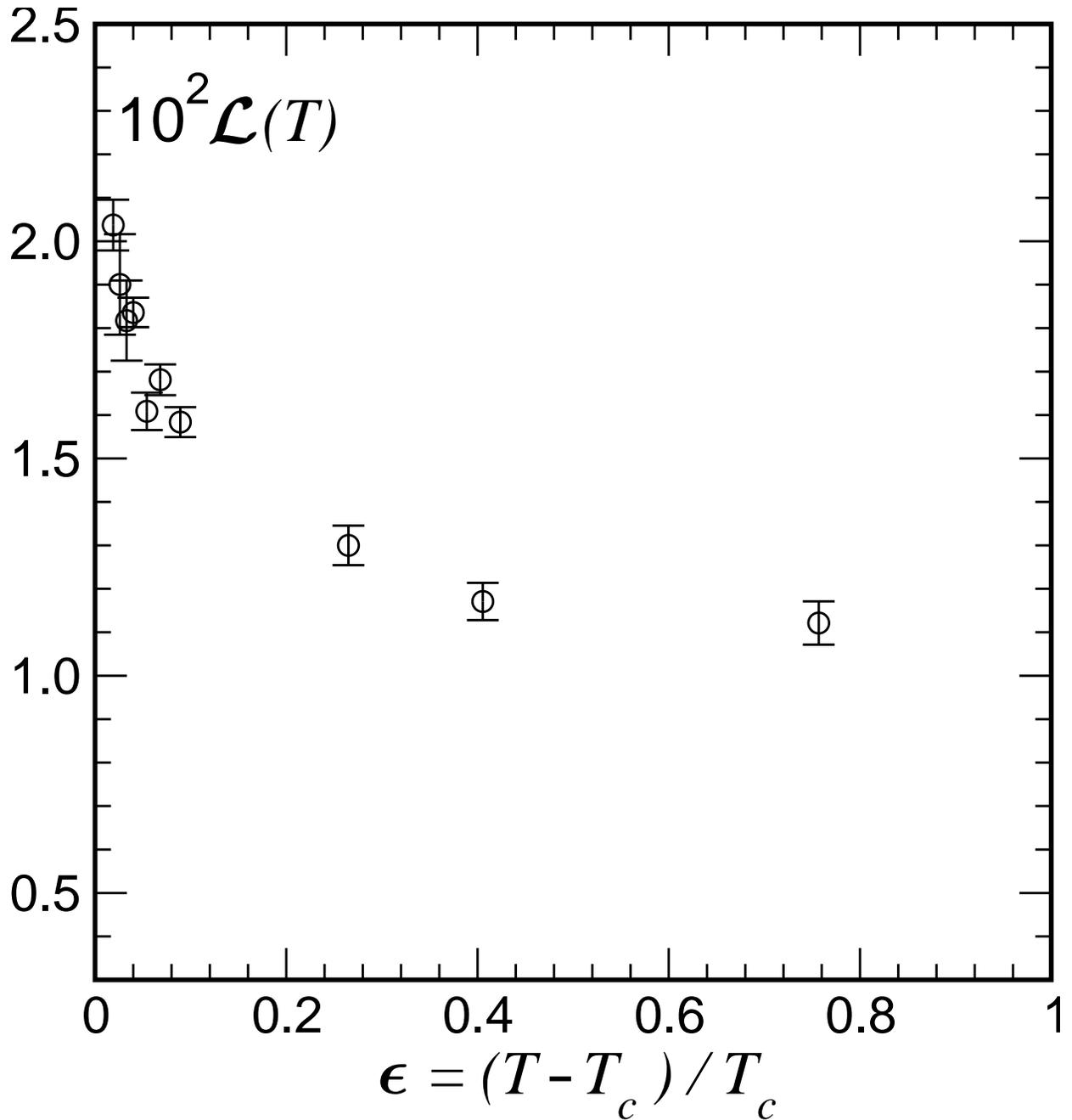}
\caption{\label{fig12} Plot of the reduced Onsager coefficient $\mathcal L(T)$
vs. $T$ for a system of $N=6400$ particles. Note the ``background''
contribution and the sharp rise as $T_c$ is approached. The four highest
data points span the range from $1.9\%$ to $4\%$ above $T_c$; but the experiments \cite{11} probe
the range $\epsilon=10^{-1}$ to $10^{-4}$.}
\end{figure}

The variation of
$\mathcal L_L(T)$ with temperature (on a linear scale) is presented in
Fig.~\ref{fig12} for the largest computationally feasible system-size
of $N=6400$ particles and, thus, of box size $L\simeq 18.6\sigma$.
Note, first, that although $\mathcal L(T)$ rises sharply close to
$T_c$, there seems to be a relatively large background contribution. If one chose
to ignore this and merely examined a direct log-log plot extending over only one decade
in $\epsilon$, the resulting effective exponent would be of little
theoretical significance. As shown by Sengers and
coworkers, \cite{15,16,xx51} the Onsager coefficient close
to criticality may be written (in the thermodynamic limit, $L\rightarrow\infty$) as
\begin{equation}\label{ons1}
\mathcal L(T) = \mathcal L_b(T) +\Delta\mathcal L(T),
\end{equation}
where $\mathcal L_b(T)$ is a slowly varying background term
which arises from fluctuations at small length scales, of order $\sigma$,
while $\Delta\mathcal L(T)$ represents the ``critical enhancement''
induced by long-range fluctuations on the scale of the diverging 
correlation length $\xi(T)\sim\epsilon^{-\nu}$. The singular piece is
predicted to diverge as
\begin{equation}\label{ons2}
\Delta\mathcal L(T)\approx QT^*/\epsilon^{\nu_{\lambda}}
\end{equation}
with an amplitude $Q$ and an exponent 
\begin{equation}\label{expnu1}
\nu_{\lambda}=x_{\lambda}\nu\simeq 0.567~~~\mbox{with}~~~x_{\lambda}=1-\eta-x_{\eta}\hspace{1mm},
\end{equation}
where $x_{\eta}$ is defined via Eq.~(\ref{singeta}) and expresses the
weak divergence of the viscosity $\eta(T)$, while $\eta\simeq 0.03$ is
the standard critical exponent \cite{42} [that enters the
scaling relation $\gamma=(2-\eta)\nu$]. 

Furthermore, we may invoke the
``extended Stokes-Einstein relation''\cite{19,xx52}
\begin{equation}\label{gnste}
\Delta D_{{\rm AB}}(T)\approx R_Dk_{{\rm B}}T/6\pi\eta(T)\xi(T),
\end{equation}
which embodies the relation $x_D=1+x_{\eta}$ [see (\ref{xdth}) and Ref.~$11$].
This is expected to describe the singular part, $\Delta D_{{\rm AB}}$, of the
mutual diffusion coefficient $D_{{\rm AB}}(T)$ and so leads to the explicit
expression
\begin{equation}\label{qq}
Q=R_D\Gamma_0\hspace{1mm}\sigma/6\pi\eta_0\xi_0
\end{equation}
for the amplitude in (\ref{ons2}). Here $R_D$ is a universal constant
of order unity while $\Gamma_0$, $\eta_0$, and
$\xi_0$ are the critical amplitudes for $\chi^*$, $\eta^*$, and $\xi$ defined
via Eqs.~(\ref{singchi}), (\ref{singeta}) and (\ref{singxi}). 

Two theoretical methods have been developed to calculate the universal dynamic amplitude
ratio $R_D$, namely, mode-coupling theory of critical dynamics and
dynamic renormalization-group theory.\cite{19} In first approximation, 
mode-coupling theory yields \cite{13} $R_D=1.00$; but when memory and nonlocal
effects are included one obtains the improved estimate \cite{xx53} $R_D=1.03$.
The early theoretical values obtained from renormalization-group theory
have varied from $0.8$ to $1.2$ due to various approximations, as reviewed by
Folk and Moser.\cite{20} The calculation of Folk and Moser with the fewest
approximations has yielded $R_D=1.063$. Experiments give values
consistent with the mode-coupling prediction.\cite{xx51,xx54} Here
we follow Luettmer-Strathmann {\it et al.}\cite{15} and adopt the estimate
$R_D=1.05$ as a compromise between the predictions of mode-coupling theory 
and the renormalization-group calculations.
Using the estimates
reported above for the amplitudes in (\ref{qq}) then yields
\begin{equation}\label{qq2}
Q=(2.8\pm0.4)\times 10^{-3}
\end{equation}
as a numerical prediction for the present model.

Our aim now is to discover if this theoretical analysis and the value (\ref{qq2})
for $Q$ are consistent with the evidence available from our
simulations. Because of the computational demands imposed by the collective
transport properties we have obtained results over a substantial 
temperature range only for the ($N=6400$)-particle system; however, for $T^*=1.48\simeq1.04~T_c^*$
we also computed $\mathcal L_L(T)$ for $N=400$, $1600$, and $3200$.

To analyze these data we write the finite-size scaling ansatz as\cite{38,41,new57}
\begin{equation}\label{fssons1}
\Delta\mathcal L_L(T)=\mathcal L(T)-\mathcal L_b(T)\approx Q\hspace{1mm}T^*W(y)/\epsilon^{\nu_{\lambda}},
\end{equation}
where $y=L/\xi$ while $W(y)$ is the finite-size scaling function. As already
discussed in the context of static critical phenomena, one requires
$W(y)\rightarrow 1$ when $y\rightarrow\infty$ so as to reproduce the
correct behavior (\ref{ons2}) in the thermodynamic limit. In this case, however, it is not clear
how rapidly $W(y)$ should approach unity. Indeed, since transport properties
are calculated from time correlation functions of currents, they reflect the nonequilibrium
behavior of the system. Although the exponentially rapid approach that applies in the static case 
[see Eq.~(\ref{scfninfty})] might still be realized here, the well known,
noncritical long-time tails in the correlation functions, etc., suggest that
a slower, power-law approach cannot be excluded.

On the other hand, for finite $L$ all properties remain bounded through
criticality so that in the limit $y\rightarrow 0$ one should have
\begin{equation}\label{scfnons}
W(y)\approx W_0y^{x_{\lambda}}[1+W_1y^{1/\nu}+...~],
\end{equation}
as in (\ref{scfnzero}), where from (\ref{expnu1}) we have 
$x_{\lambda}\simeq 0.90$.

Of course, we do not know the value of the background $\mathcal L_b(T)$
in (\ref{fssons1}); but since it is slowly varying, we may reasonably replace 
it by a constant effective value $\mathcal L_b^{eff}$. Then, by treating $\mathcal L_b^{eff}$ as an adjustable
parameter and examining the simulation data for 
$\mathcal W_L(T)=\Delta\mathcal L_L(T)\epsilon^{\nu_{\lambda}}/T^*$
as $T$ and $L$ vary with $\nu_{\lambda}$ set to its Ising value, we may seek an 
optimal data collapse onto the scaling form $QW[L/\xi(T)]$. Note
that if this is achieved, the value $Q$ should emerge when $y=L/\xi$ becomes
large.\cite{new57}

\begin{figure}[htb]
\centering
\includegraphics*[width=0.9\textwidth]{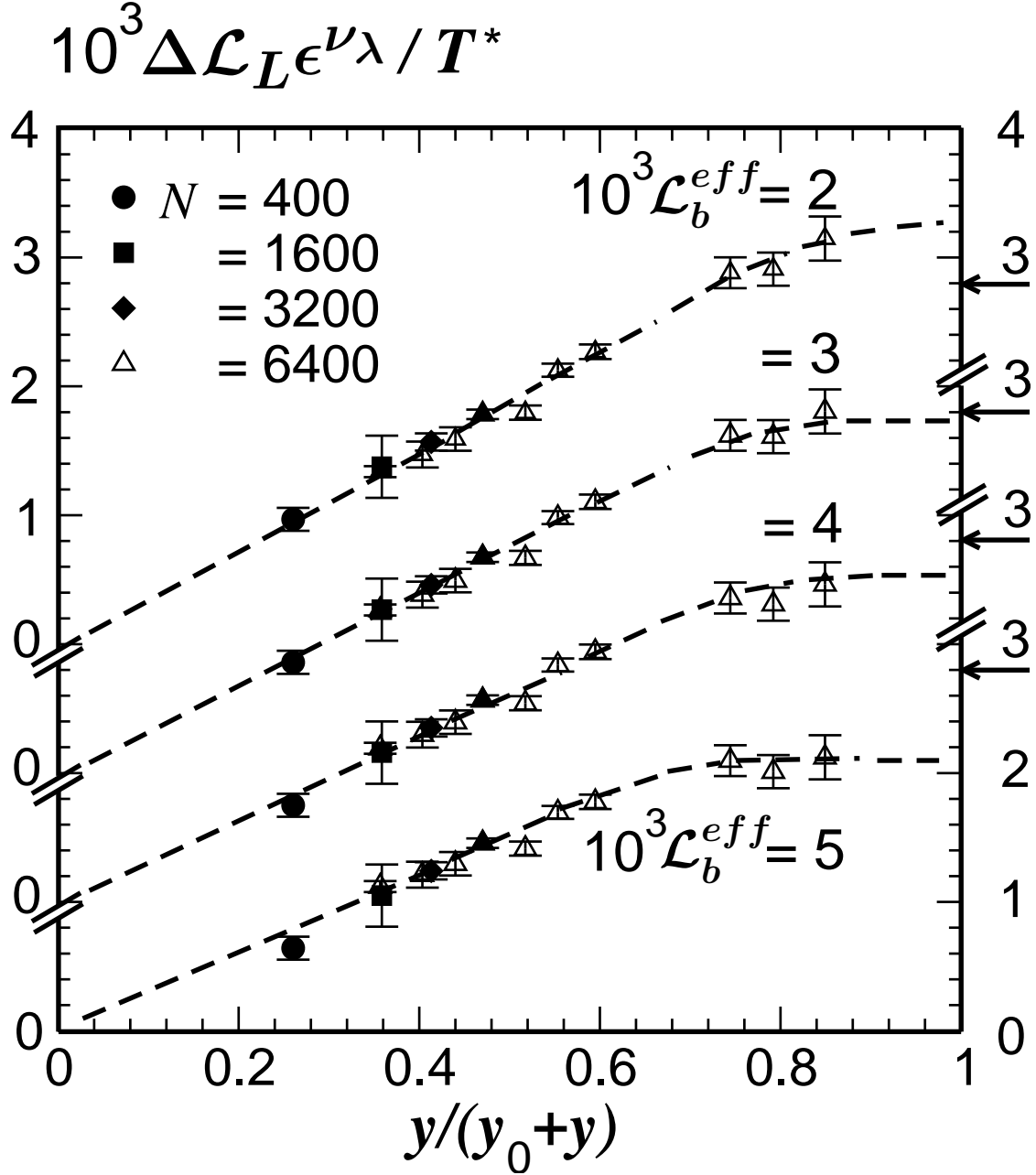}
\caption{\label{fig13}
Finite-size scaling plots for the interdiffusional Onsager coefficient
$\mathcal L_L(T)$ with $\epsilon=(T-T_c)/T_c$, $y=L/\xi(T)$, and trial
values for the effective background contribution $\mathcal L_b^{eff}$. The approximate Ising
value $\nu_{\lambda}=0.567$ has been adopted and, for convenience, we have set $y_0=7$ in
the abscissa variable, $y/(y_0+y)$, that approaches unity when $L\rightarrow\infty$.
The filled symbols represent data at $\epsilon\simeq 4.0\times 10^{-2}$ for different system sizes
of $N=400$ to $6400$ particles and fixed density
$\rho\sigma^3=1$. The solid arrows on the right hand axis indicate the
central theoretical estimate for the critical amplitude $Q$: see text.}
\end{figure}

Fig.~\ref{fig13} presents separated plots of $\mathcal W_L(T)$ vs. $y/(y+y_0)$, with, as
in Fig~\ref{fig3}, $y_0=7$, for four assignments of $\mathcal L_b^{eff}$.
Note that the filled symbols represent the data at $T^*=1.48$ for system sizes
$L/\sigma\simeq 7.37,~11.70,~14.74$, and $18.57$; their reasonably good
collapse onto the remaining data (all for $L/\sigma\simeq 18.57$) and their approach
towards $0$ for small $y$ serve to justify
\begin{equation}\label{lbcoll}
\mathcal L_b^{eff}=(3.3\pm0.8)\times 10^{-3}
\end{equation}
as a sensible estimate of the background term in the Onsager coefficient:
compare with Fig.~\ref{fig12}. The horizontal arrows marked on the right side of
Fig.~\ref{fig13} indicate the central theoretical value (\ref{qq2}) for the
amplitude $Q$. It is evident that the agreement is surprisingly good. Indeed, 
had one been asked to estimate $Q$ from these plots one might have proposed 
$Q=(2.7\pm0.4)$$\times$$10^{-3}$, again surprisingly close to the theoretical value.
Further details of this finite-size scaling analysis, including a fit for
$\mathcal W_L(T)$, are presented in Ref. $58$.

Thus we conclude that our simulation data are, in fact, fully consistent
with the predictions of the theory including the value $0.567$ for the exponent $\nu_{\lambda}$,
and, hence, the result $x_D\simeq 1.0679$ for the interdiffusion coefficient
itself. It cannot be emphasized too strongly, however, that our discussion
demonstrates that in the analysis of
simulations near critical points one needs to account properly for the inevitable
finite-size effects and, when theory indicates, also for appropriate background
contributions typically arising from short-range fluctuations.\cite{new57}

\section{SUMMARY}

We have studied the static and dynamic properties of a symmetric truncated
Lennard-Jones binary fluid model with
$\sigma_{{\rm AA}}=\sigma_{{\rm BB}}=\sigma_{{\rm AB}}=\sigma$,
$\varepsilon_{{\rm AA}}=\varepsilon_{{\rm BB}}=2\varepsilon_{{\rm AB}}=\varepsilon$ and
masses $m_{\rm A}=m_{\rm B}=m$. This model has a liquid-liquid miscibility gap. We have used
a combination of semi-grandcanonical Monte Carlo (SGMC) and
microcanonical molecular dynamics simulations to study both the
static and dynamic properties near the demixing critical point.
The symmetry of the model sets the critical composition at
$x_{\rm A}=x_{\rm B}=1/2$. We have studied the system at the comparatively high liquid density 
$\rho^*=\rho\sigma^3=1$, in which region 
the gas-liquid and liquid-solid transitions are far from the temperature
range of interest.

The critical temperature $T_c$ has been determined quite accurately as
$T_c^*\equiv k_{{\rm B}}T_c/\varepsilon=1.4230\pm0.0005$ using a variety of techniques. 
Because of the short-range
nature of the interactions one anticipates that demixing criticality in the model belongs to
the three-dimensional Ising universality class. All our data for the
static properties near the critical point strongly support that presumption.

We have also presented the first comprehensive study of the dynamic properties
of a binary fluid near the critical point. We find evidence for a very weak divergence 
of the shear viscosity, $\eta(T)$, near the critical point 
in accord with expectations. 
The self-diffusion constant $D(T)$
remains finite at the critical point which is consistent with
some earlier studies. We also find that the Stokes-Einstein relation
remains a fairly good approximation even within $0.5\%$ of $T_c$.

In contrast to the self-diffusion constant, the
interdiffusion constant $D_{{\rm AB}}(T)$ vanishes 
rapidly when $T\rightarrow T_c$. Our analysis
of the simulation data supports the various theoretical predictions for the
critical exponents of all these quantities including the 
dynamic exponent relation \cite{11,19,xx52} $x_D=1+x_{\eta}$. But, 
even with an
accurate knowledge of $T_c$ and of the correlation length and concentration susceptibility,
it proves essential to
consider the finite-size effects and allow for background
contributions arising from short-range fluctuations, in order to 
properly analyze the data for the interdiffusion coefficient. 

Finally, however, we have not discussed the bulk viscosity,
$\eta_B(T)$, which is expected to diverge much more rapidly
than the shear viscosity.\cite{21} That remains a significant task for future work.

\section*{Acknowledgement} M.E.F. and S.K.D. are grateful for support from
the National Science Foundation under Grant No.~CHE $03$-$01101$. S.K.D. also acknowledges
financial support from the Deutsche Forschungsgemeinschaft (DFG) via Grant No.~Bi $314/18$-$2$
and thanks Professor Kurt Binder and Dr. J\"urgen Horbach
for supporting his stay in the Johannes
Gutenberg Universit\"at Mainz, Germany, where all the simulations were
carried out with their close collaboration.

\newpage

\clearpage

\clearpage

\clearpage

\clearpage

\clearpage

\clearpage

\clearpage

\clearpage

\clearpage

\clearpage

\clearpage

\clearpage

\clearpage


\begin{thebibliography}{99}
\bibitem{yy42} Depending upon the proximity to the critical point, average
was taken over 
$10$ to $15$ distinct initial configurations.
\bibitem{xx45} Note that the calculation of the mean square displacements was
not subject to periodic boundary conditions, so they can grow without limit.
\end{thebibliography}
\end{document}